\documentclass[12pt,preprint]{aastex}

\slugcomment{}

\shorttitle{The Fermi high Latitude Survey}
\shortauthors{Ajello et al.}

\begin{document}


\title{The Fermi-LAT high-latitude Survey: Source Count Distributions and
the Origin of the Extragalactic Diffuse Background} 

\author{
A.~A.~Abdo\altaffilmark{2,3}, 
M.~Ackermann\altaffilmark{4}, 
M.~Ajello\altaffilmark{4,1}, 
E.~Antolini\altaffilmark{5,6}, 
L.~Baldini\altaffilmark{7}, 
J.~Ballet\altaffilmark{8}, 
G.~Barbiellini\altaffilmark{9,10}, 
D.~Bastieri\altaffilmark{11,12}, 
B.~M.~Baughman\altaffilmark{13}, 
K.~Bechtol\altaffilmark{4}
% ...
}
\altaffiltext{1}{Corresponding authors: M.~Ajello, majello@slac.stanford.edu; A.~Tramacere, tramacer@slac.stanford.edu.}
\altaffiltext{2}{Space Science Division, Naval Research Laboratory, Washington, DC 20375, USA}
\altaffiltext{3}{National Research Council Research Associate, National Academy of Sciences, Washington, DC 20001, USA}
\altaffiltext{4}{W. W. Hansen Experimental Physics Laboratory, Kavli Institute for Particle Astrophysics and Cosmology, Department of Physics and SLAC National Accelerator Laboratory, Stanford University, Stanford, CA 94305, USA}
\altaffiltext{5}{Istituto Nazionale di Fisica Nucleare, Sezione di Perugia, I-06123 Perugia, Italy}
\altaffiltext{6}{Dipartimento di Fisica, Universit\`a degli Studi di Perugia, I-06123 Perugia, Italy}
\altaffiltext{7}{Istituto Nazionale di Fisica Nucleare, Sezione di Pisa, I-56127 Pisa, Italy}
\altaffiltext{8}{Laboratoire AIM, CEA-IRFU/CNRS/Universit\'e Paris Diderot, Service d'Astrophysique, CEA Saclay, 91191 Gif sur Yvette, France}
\altaffiltext{9}{Istituto Nazionale di Fisica Nucleare, Sezione di Trieste, I-34127 Trieste, Italy}
\altaffiltext{10}{Dipartimento di Fisica, Universit\`a di Trieste, I-34127 Trieste, Italy}
\altaffiltext{11}{Istituto Nazionale di Fisica Nucleare, Sezione di Padova, I-35131 Padova, Italy}
\altaffiltext{12}{Dipartimento di Fisica ``G. Galilei", Universit\`a di Padova, I-35131 Padova, Italy}
\altaffiltext{13}{Department of Physics, Center for Cosmology and Astro-Particle Physics, The Ohio State University, Columbus, OH 43210, USA}
% ...

\email{majello@slac.stanford.edu}

%
%
%
%
%
%
%
%

%
%

\begin{abstract}
This is the first of a series of papers aimed at characterizing the
populations detected in the high-latitude sky of the {\it Fermi}-LAT survey.
In this work we focus on the intrinsic spectral and flux properties
of the source sample. We show that when selection effects
are properly taken into account,  {\it Fermi} sources are on average
steeper than previously found (e.g. in the bright source list)
 with an average photon index of 2.40$\pm0.02$
over the entire 0.1--100\,GeV energy band. We confirm that FSRQs have 
steeper spectra
than BL Lac objects with an average index of 2.48$\pm0.02$ versus 
2.18$\pm0.02$. Using several methods we build the deepest source
count distribution at GeV energies
deriving that the intrinsic source (i.e. blazar)
surface density at F$_{100}\geq10^{-9}$\,ph cm$^{-2}$ s$^{-1}$
is  0.12$^{+0.03}_{-0.02}$\,deg$^{-2}$.
The integration of the source count distribution
yields that point sources contribute 16$(\pm1.8)$\,\% 
($\pm$7\,\% systematic uncertainty)
of the GeV isotropic diffuse background. 
At the fluxes currently reached by LAT
we can rule out the hypothesis
that point-like sources (i.e. blazars) produce a larger
fraction of the diffuse emission.\\
\end{abstract}


\section{Introduction}

The origin of the extragalactic gamma-ray  background (EGB) at GeV $\gamma$-rays
is one of the fundamental unsolved problems in astrophysics. 
The EGB was first detected by the SAS-2 mission
\citep{fichtel95} and its  spectrum was measured with good accuracy by 
the Energetic Gamma Ray Experiment Telescope  
\citep[EGRET][]{sreekumar98,strong04} on board the Compton Observatory.
These observations by themselves do not provide much insight into the
sources of the EGB.

Blazars, active galactic nuclei
(AGN) with a relativistic jet pointing close to our line of sight,
represent the most numerous
population detected by EGRET \cite{hartman99}
and their flux constitutes 15\,\% of the total EGB intensity 
(resolved sources plus diffuse emission). Therefore,
undetected blazars (e.g. all the blazars
under the sensitivity level of EGRET) are the most likely
candidates for the origin of the bulk of  the EGB emission. 
Studies of the luminosity function of blazars showed that
the contribution of blazars to the EGRET EGB could be in the range
from 20\,\% to 100\,\% \citep[e.g.][]{stecker96,chiang98,muecke00}, although
the newest derivations suggest that 
blazars are responsible for only $\sim20$--$40$\,\%
of the EGB \citep[e.g.][]{narumoto06,dermer07,inoue09}.

It is thus possible that the EGB emission encrypts in itself the signature
of some of the most powerful and interesting phenomena in astrophysics.
Intergalactic shocks produced by the assembly of Large Scale Structures
\citep[e.g.][]{loeb00,miniati02,keshet03,gabici03}, $\gamma$-ray
emission from galaxy clusters  \citep[e.g.][]{berrington03,pfrommer08}, 
emission from starburst as well as normal 
galaxies  \citep[e.g.][]{pavlidou02,thompson07},  are among
the most likely candidates for the generation of diffuse the GeV emission.
Dark matter (DM) which constitutes more than 80\,\% of the matter in
the Universe can also provide a diffuse, cosmological, background
of $\gamma$-rays. Indeed,  supersymmetric theories with R-parity
predict that the lightest DM particles
(i.e., the neutralinos) are stable and can annihilate into GeV
$\gamma$-rays \citep[e.g.][]{jungman96,bergstrom00,ullio02,ahn07}.

With the advent of the {\it Fermi} Large Area Telescope (LAT) a better
understanding of the origin of the GeV diffuse emission becomes possible.
{\it Fermi} has recently performed a new measurement of the EGB spectrum
\citep[also called isotropic diffuse background,][]{lat_edb}. This
has been found to be consistent with a featureless power law with
a photon index of $\sim$2.4 in the 0.2--100\,GeV energy range.
The integrated flux (E$\geq$100\,MeV) of 
1.03$(\pm0.17)\times10^{-5}$\,ph cm$^{-2}$ s$^{-1}$ sr$^{-1}$ has been
found to be significantly lower than the one of
 1.45($\pm0.05$)$\times10^{-5}$\,ph cm$^{-2}$ s$^{-1}$ sr$^{-1}$ determined from EGRET data \citep[see][]{sreekumar98}.

In this study we address the contribution of {\it unresolved} point sources 
to the GeV diffuse emission and we discuss the implications.
Early findings on the integrated emission 
of {\it unresolved} blazars were already reported in \cite{lat_lbas} using
a sample of bright AGN detected in the first three months of {\it Fermi}
observations.
The present work represents a large advance, with $\sim$4 times more blazars and a detailed investigation of selection effects in source detection.

This work is organized as follows. In $\S$~\ref{sec:spec} the
intrinsic spectral properties of the {\it Fermi} sources are determined.
In $\S$~\ref{sec:sim} the Monte Carlo simulations used for
this analyses are outlined with the inherent systematic
uncertainties (see $\S$~\ref{sec:syst}). Finally the source
counts distributions are derived in $\S$~\ref{sec:logn} and 
$\S$~\ref{sec:bands} while
the contribution of point sources to the GeV diffuse background
is determined in $\S$~\ref{sec:edb}. $\S$~\ref{sec:discussion}
discusses and summarizes our findings.
Since the final goal of this work is deriving the contribution of
sources to the EGB, we will only use physical quantities (i.e. source flux
and photon index) averaged over the time (11 months) included in the 
analysis for the First {\it Fermi}-LAT catalog \citep[1FGL,][]{cat1}.

\section{Terminology}
\label{sec:term}
Throughout this paper we use a few terms which might not be familiar
to the reader. In this section meanings of the most often
used are clarified.
\begin{itemize}
\item {\it spectral bias}: (or photon index bias) is the 
selection effect which allows {\it Fermi}-LAT to detect spectrally 
hard sources at fluxes generally fainter than for soft sources.
\item {\it flux-limited} sample: it refers to a sample which is 
selected uniformly according solely to the source flux. If the
flux limit is chosen to be bright enough (as in the case of this paper), 
then the selection effects  affecting any other properties
(e.g. the source spectrum) of the sample are negligible. 
This is a truly uniformly selected sample.
\item {\it diffuse emission from unresolved point sources}: 
it represents a measurement of the integrated emission from sources
which have not been detected by {\it Fermi}.
As it will be shown in the next sections, for each source detected at low
fluxes, there is a large number of sources which have not been detected because
of selection effects (e.g. the local background was too large 
or the photon index was too soft, or a combination of both). 
The diffuse emission from {\it unresolved} point
sources (computed in this work)
addresses  the contribution of 
all those sources which have not been detected because of these
selection effects, 
but have a flux which is formally larger than the faintest {\it detected}
source.
%

\end{itemize}

\section{Average Spectral Properties}
\label{sec:spec}

\subsection{Intrinsic Photon index distributions}
\label{sec:photon}
\begin{figure*}[ht!]
  \begin{center}
  \begin{tabular}{cc}
    \includegraphics[scale=0.4]{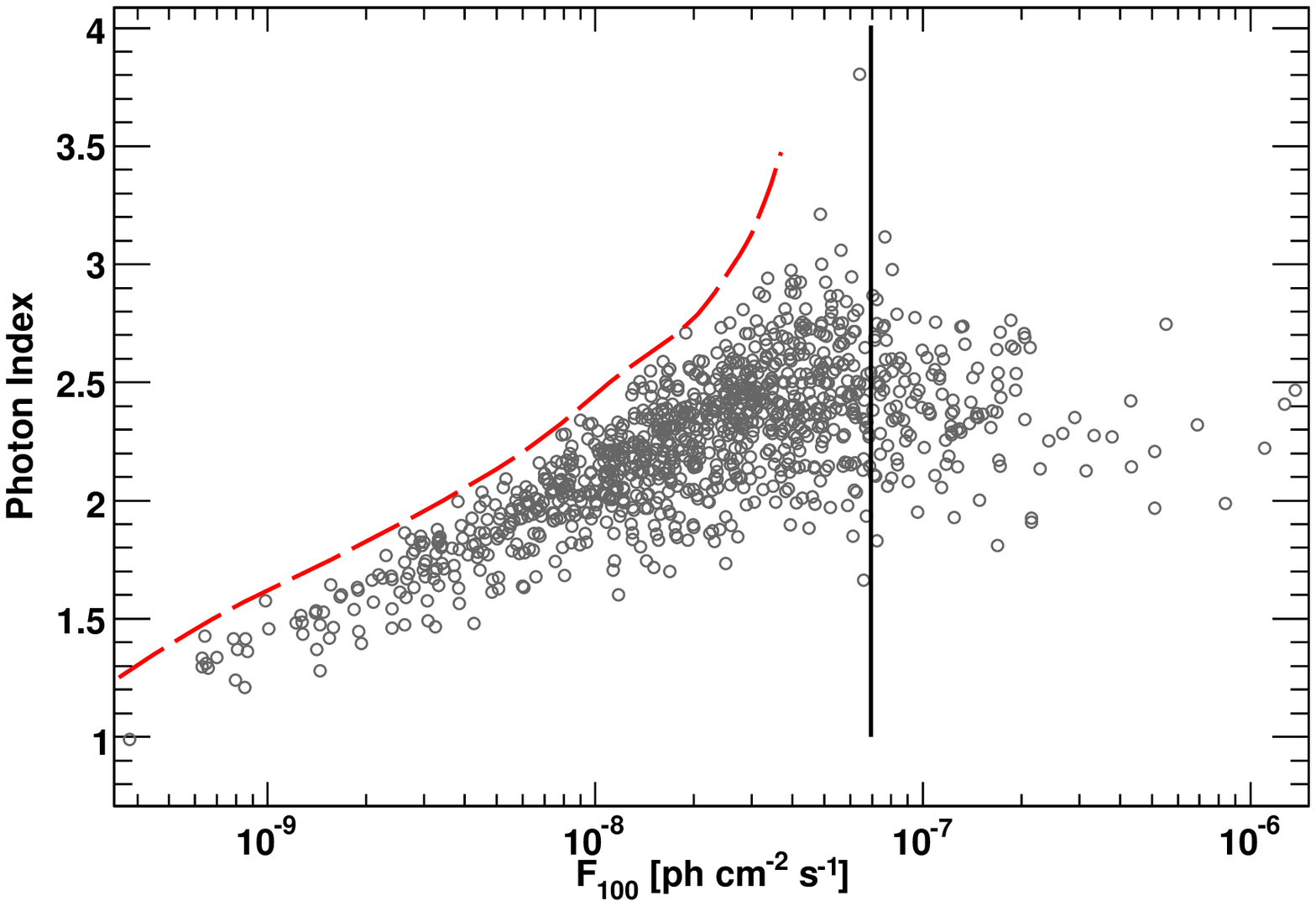} 
	 \includegraphics[scale=0.4]{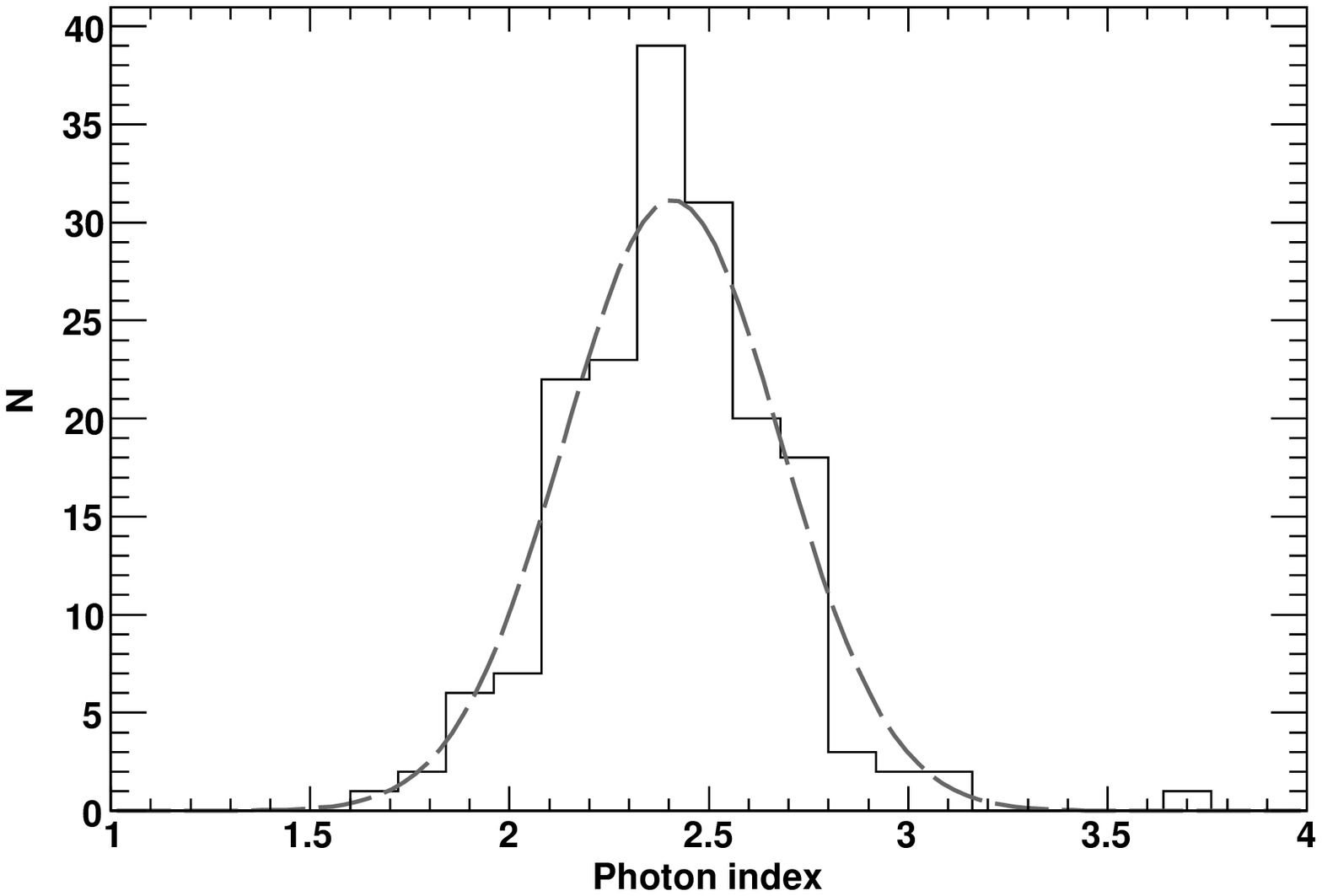}\\
\end{tabular}
  \end{center}
  \caption{
{\bf Left Panel:}Flux-photon index plane for all the $|b|\geq10^{\circ}$
sources with TS$\geq25$. The dashed line is the flux limit
as a function of photon index 
reported in \cite{agn_cat}, while the solid line represents the limiting
flux above which the spectral selection effects become negligible.
{\bf Right Panel:}
Photon index distribution of all sources for
 F$_{100}\geq7\times 10^{-8}$\,ph cm$^{-2}$ s$^{-1}$. Above
this limit the LAT selection effect towards hard sources becomes
negligible.
}
  \label{fig:idx_f}
\end{figure*}

As shown already in \cite[][but see also Fig.~\ref{fig:idx_f}]{lat_lbas}, 
at faint fluxes the LAT detects more easily hard-spectrum sources
rather than sources with a soft spectrum. Sources with a photon index 
(e.g. the exponent of the power-law fit to the source photon spectrum)
of  1.5 can be detected to fluxes which are a factor $>20$ fainter
than those at which a source with a photon index of 3.0 can be detected
\citep[see][for details]{agn_cat}. Thus, given this
 strong selection effect,
the intrinsic photon index distribution is necessarily different 
from the observed one.
An approach to recover the intrinsic photon index distribution is
obtained by studying the sample above 
F$_{100}\approx 7\times 10^{-8}$\,ph cm$^{-2}$ s$^{-1}$ and 
$|b|\geq10^{\circ}$ (see right panel of   Fig.~\ref{fig:idx_f}).
Indeed above this flux limit, LAT detects all sources irrespective
of their photon index, flux or position in the high-latitude sky.
Above this limit LAT detects 135 sources.
Their photon index distribution, reported in Fig.~\ref{fig:idx_f} 
is compatible with a Gaussian
distribution with mean of 2.40$\pm0.02$ and dispersion of 0.24$\pm0.02$.
These values differ from the mean of 2.23$\pm0.01$ and dispersion
of 0.33$\pm0.01$ derived using the entire $|b|\geq10^{\circ}$ sample.
Similarly the intrinsic photon-index distributions of FSRQs and BL Lacs
are different from the observed distributions. In both case the {\it observed
} average photon-index is harder than the intrinsic average value.
The results are summarized in Tab.~\ref{tab:index}.

\begin{deluxetable}{lccccc}
\tablecolumns{6} 
\tablewidth{0pc}
\tabletypesize{\footnotesize}
\tablecaption{
Observed versus Intrinsic photon indices distributions for the 
Fermi/LAT source classes.
\label{tab:index}}
\tablehead{
\colhead{}    &  
\multicolumn{2}{c}{Observed Distribution}  &   \colhead{}    &
\multicolumn{2}{c}{Intrinsic Distribution}\\
\cline{2-3} \cline{5-6} \\ 
\colhead{SAMPLE} & 
\colhead{mean}   & \colhead{$\sigma$}  & \colhead{} &
\colhead{mean}   & \colhead{$\sigma$} 
}
\startdata
ALL    & 2.24$\pm0.01$ &0.31$\pm0.01$  & & 2.40$\pm0.02$ & 0.24$\pm0.02$ \\
FSRQ   & 2.44$\pm0.01$ & 0.21$\pm0.01$ & & 2.47$\pm0.02$ & 0.19$\pm0.02$ \\
BL Lac & 2.05$\pm0.02$ & 0.29$\pm0.01$ & & 2.20$\pm0.04$ & 0.22$\pm0.03$\\

\enddata
\end{deluxetable}
\subsection{Stacking Analysis}
\label{sec:stacking}
\begin{figure}[h!]
\begin{centering}
	\includegraphics[scale=0.6]{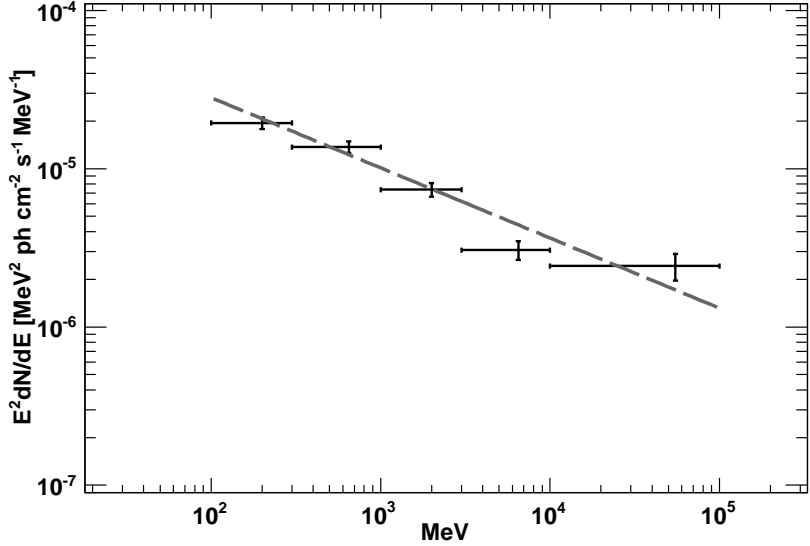} 
	\caption{Stacked spectrum of sources in the {\it flux-limited} sample.
The dashed line is the best power law fit with a slope of 2.41$\pm0.07$.
	\label{fig:stack}}
\end{centering}
\end{figure}

Another way to determine the average spectral properties is by stacking
source spectra together. This is particularly simple since 
\cite[][]{cat1} reports the source flux in five different 
 energy bands. We thus performed a stacking analysis
of those sources with  F$_{100}\geq7\times 10^{-8}$\,ph cm$^{-2}$ s$^{-1}$,
TS$\geq25$, and $|b|\geq$10$^{\circ}$. For each energy band the average
flux is computed as the weighted average of all source fluxes in that
band using the inverse of the flux variance as a weight.
The average spectrum is shown in Fig.~\ref{fig:stack}. A 
 power law model gives a satisfactory fit to the data (e.g.
$\chi^2/dof\approx 1$), yielding
a photon index of 2.41$\pm0.07$ in agreement with the results
of the previous section.

We repeated the same exercise separately for sources identified as
FSRQs and BL Lacs in the {\it flux-limited} sample.
Both classes have an average spectrum which is compatible
with a single power law over the whole energy band.
FSRQs are characterized by an index of 2.45$\pm0.03$ while BL Lac objects have
an average index of 2.23$\pm0.03$

\begin{figure*}[ht!]
  \begin{center}
  \begin{tabular}{cc}
    \includegraphics[scale=0.4]{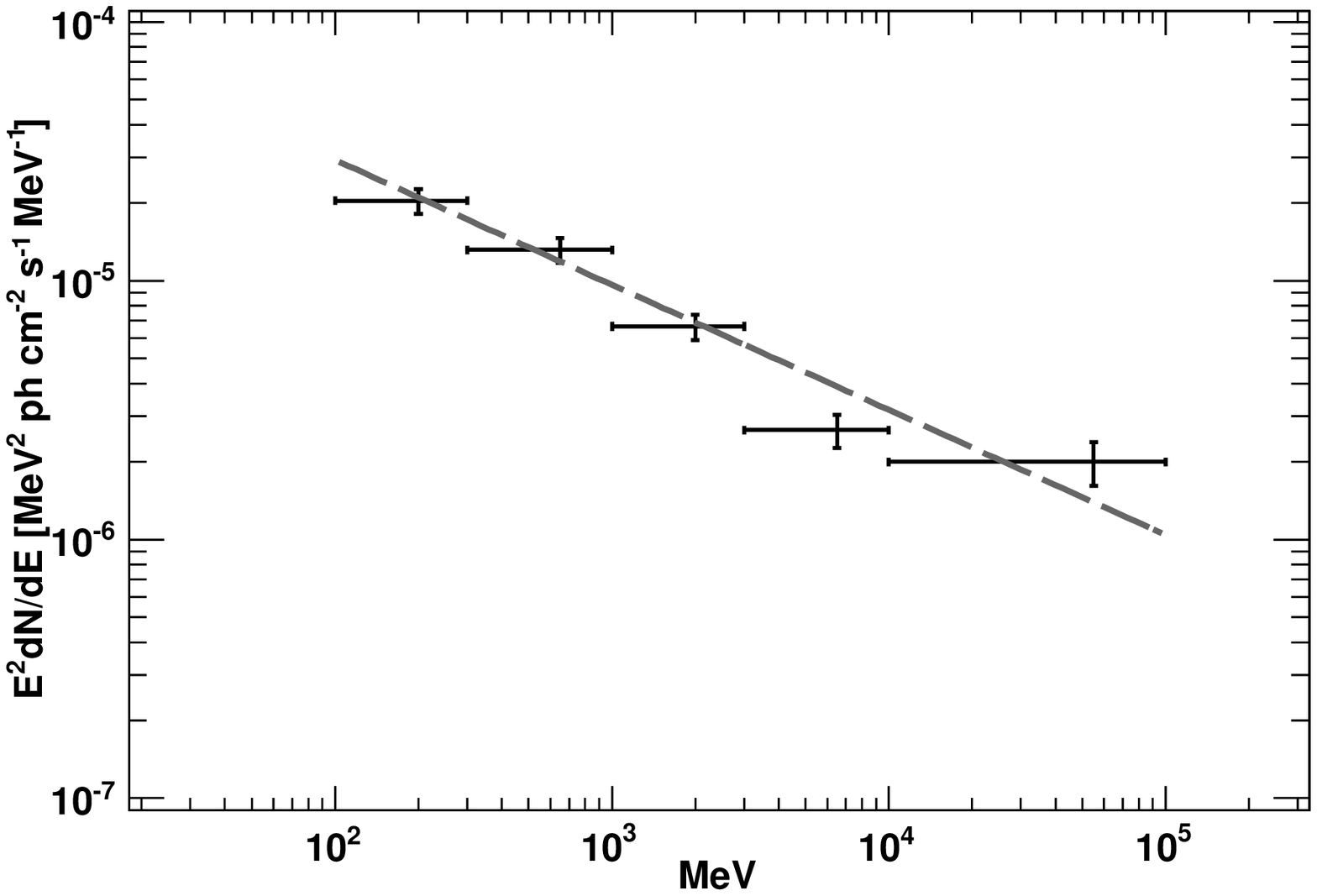} 
	 \includegraphics[scale=0.4]{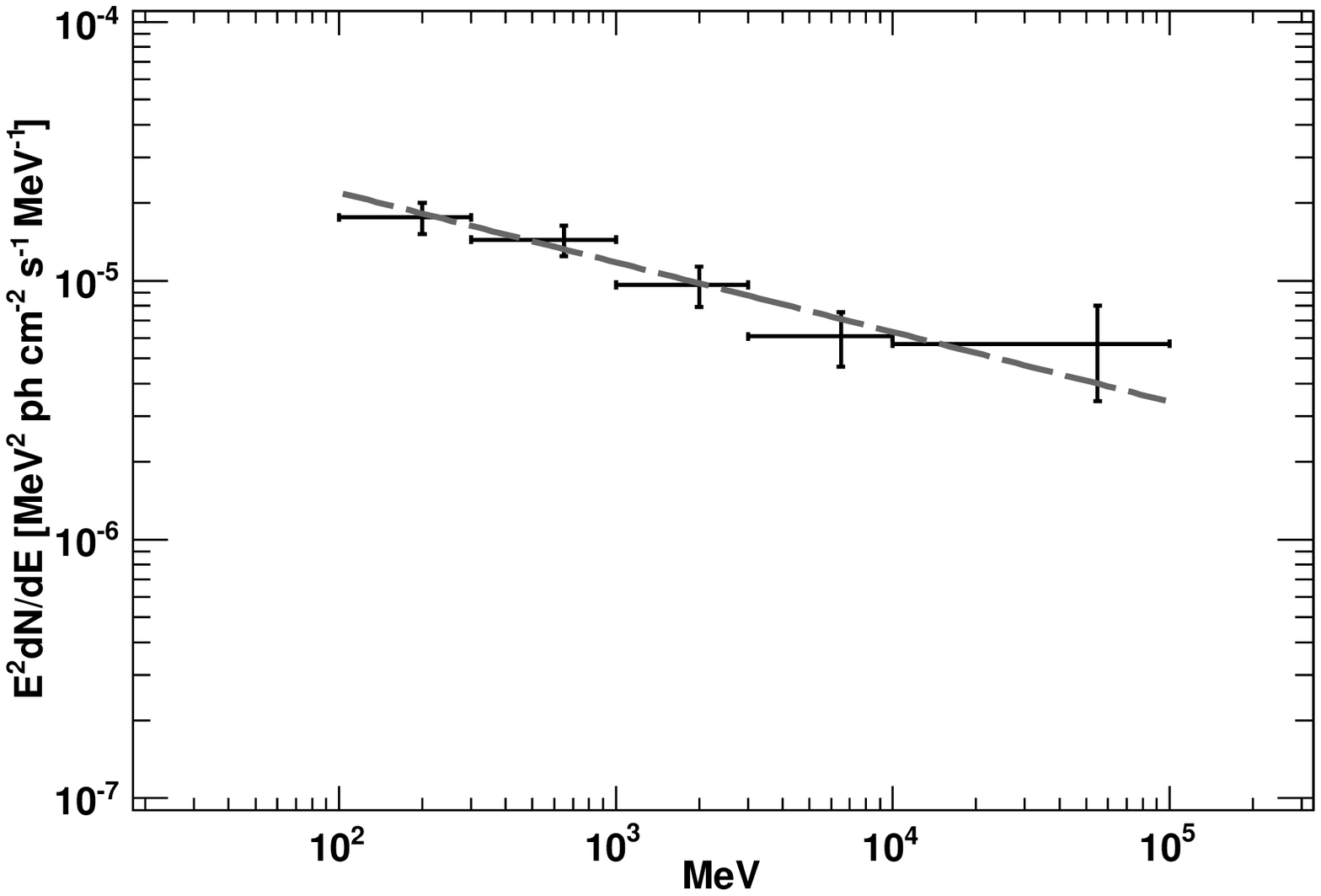}\\
\end{tabular}
  \end{center}
  \caption{
Stacked spectrum of FSRQs (left) and BL Lac objects (right) in the {\it Fermi}-LAT
{\it flux-limited} sample.
}
  \label{fig:stack_class}
\end{figure*}

\section{Monte Carlo Simulations}
\label{sec:sim}

In order to estimate the LAT sky coverage robustly 
we performed detailed Monte Carlo simulations.
The scheme of the simulation procedure is an improved version
of what has already been applied in \cite{lat_lbas}.
We performed 18 end-to-end simulations of the LAT sky which
resemble as closely as possible the observed one.
The tool {\it gtobssim}\footnote{The list of science tools for the analysis
of {\it Fermi} data is accessible at http://fermi.gsfc.nasa.gov/ssc/data/analysis/scitools/overview.html.} has been used for this purpose.
For each simulation we  modeled the Galactic and isotropic diffuse backgrounds
using models (e.g. gll\_iem\_v02.fit) 
currently recommended by the LAT team. 

An isotropic  population of point-like sources 
was added to each simulated observation. 
The coordinates of each source were randomly drawn in order
to produce an isotropic distribution on the sky. Source fluxes were
randomly drawn from a standard log $N$--log $S$ distribution with parameters
similar to the one observed by LAT (see next section). Even though the method
we adopt to derive the survey sensitivity does not depend on the 
normalization or the slope of the input log $N$--log $S$, using the real
distribution allows simulated observations to be produced that closely 
resemble the sky as observed with the LAT.
The photon index of each source was also drawn from a Gaussian
distribution with mean of 2.40 and 1\,$\sigma$ width of 0.28. As noted in
the previous section, this distribution represents well the intrinsic
(not the observed one) distribution of photon indices. The adopted dispersion
is slightly larger than what was found in the previous section and it
is derived from the analysis of the entire sample (see $\S$~\ref{sec:logn_2d}). 
In this framework we are neglecting any possible dependence of the photon
index distribution with flux.
Also we remark that the approach used here to derive the source count
distribution depends very weakly on the assumptions (e.g.
the log $N$--log $S$ used) made in the simulations.

More than 45000 randomly distributed sources have been generated for each
realization of the simulations. Detection follows (albeit in a simpler
way) the scheme used in \cite{cat1}.
This scheme adopts three energy bands for source detection.
The first band includes all {\it front}-converting\footnotemark{} and 
{\it back}-converting photons with 
energies larger than 200\,MeV and 400\,MeV, respectively.
\footnotetext{Photons pair-converting in the top 12 layers of the tracker
are classified as {\it front-}converting photons or {\it back-}converting otherwise. }
The second band starts at 1\,GeV for {\it front} photons and at 2\,GeV
for {\it back} photons. The high-energy band starts at 5\,GeV
for {\it front} photons and at 10\,GeV for {\it back} photons.
The choice of combining  {\it front} and {\it back} events
with different energies
is motivated by the fact that {\it  front} events have
a better point spread function (PSF) than {\it back} ones.
The two PSFs are similar when the energy of {\it back}-converting
photons is approximately twice as that of {\it front}-converting ones. 
The image pixel sizes changes according to the energy band
and is 0.1, 0.2 and 0.3 degrees for the low, medium and high-energy
bands respectively. The final list of {\it candidate} sources
is obtained starting the detection at the highest energy band
and adding all those sources which, being detected  at lower energy,
have a position not consistent with those detected at high energy.
The detection step uses {\it pgwave} for determining the position
of the excesses and {\it pointfit} for refining the source position.
{\it Pgwave} \citep{ciprini07}
is a tool which uses several approaches (e.g. wavelets,
thresholding, image denoising and a sliding cell algorithm) to
find source candidates while {\it pointfit} \citep[e.g.][]{burnett09}
employes a simplified binned likelihood algorithm
to optimize the source position.

All the source candidates found at this stage are then ingested to the
Maximum Likelihood (ML) algorithm {\it gtlike} to determine the 
significance and the spectral parameters. In this step all sources'
spectra are modeled as single power laws.
On average, for each simulation only $\sim$1000 sources are detected (out
of the 45000 simulated ones)
above a TS\footnote{The test statistics (or TS) is defined as:
TS=$-2({\rm ln} L_0 - {\rm ln} L_1)$. Where $L_0$ and $L_1$
are the likelihoods of the background (null hypothesis) and 
the hypothesis being tested (e.g. source plus background). 
According to \cite{wilks38}, the significance of a detection
is approximately $n_{\sigma}=\sqrt(TS)$ \citep[see also][]{ajello08a}.} 
of 25 and this is found to be  in good agreement with the real data.

\subsection{Performances of the detection algorithm on real data}
\label{sec:cat}
\begin{figure*}[ht!]
  \begin{center}
  \begin{tabular}{cc}
    \includegraphics[scale=0.4]{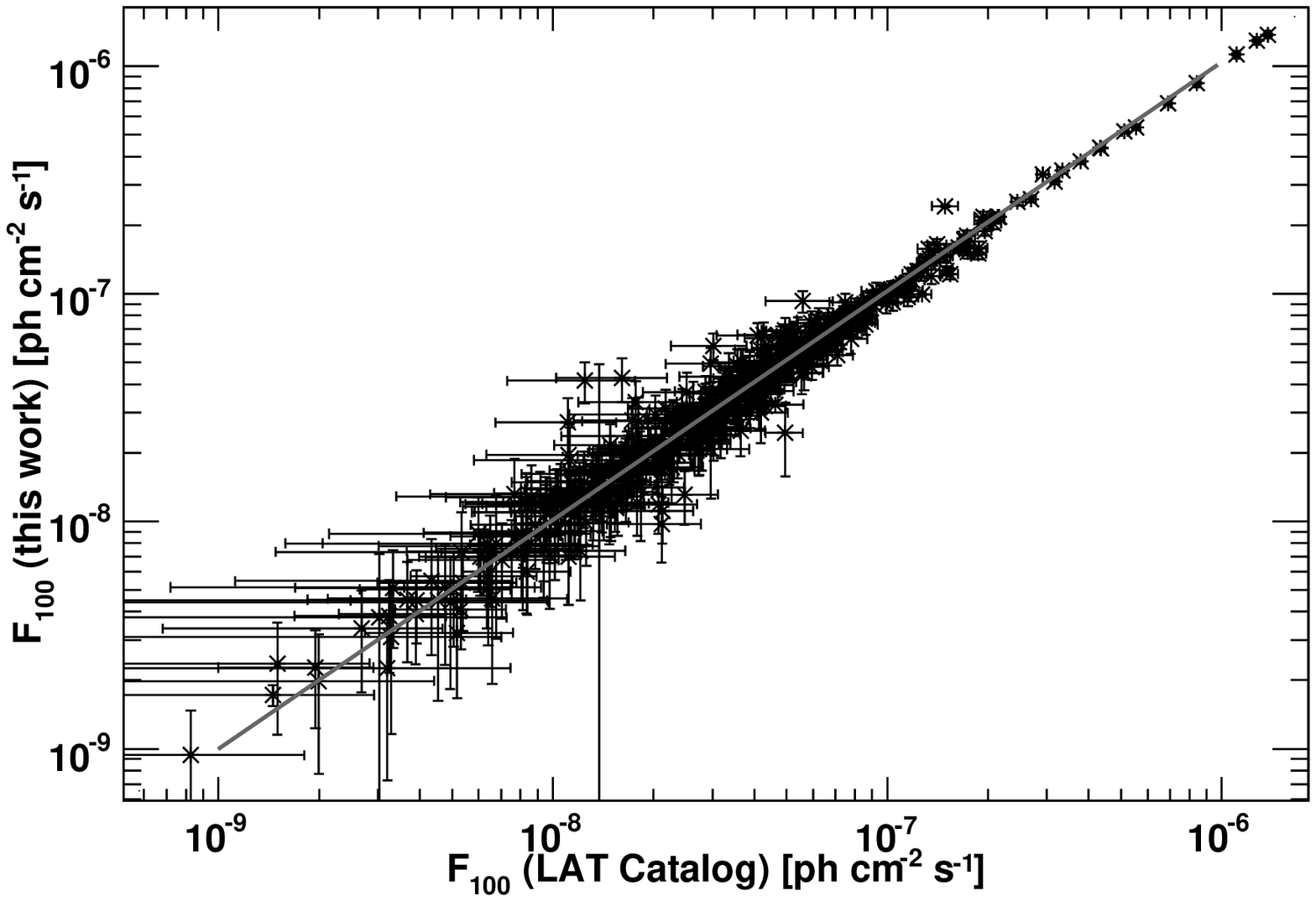} 
	 \includegraphics[scale=0.4]{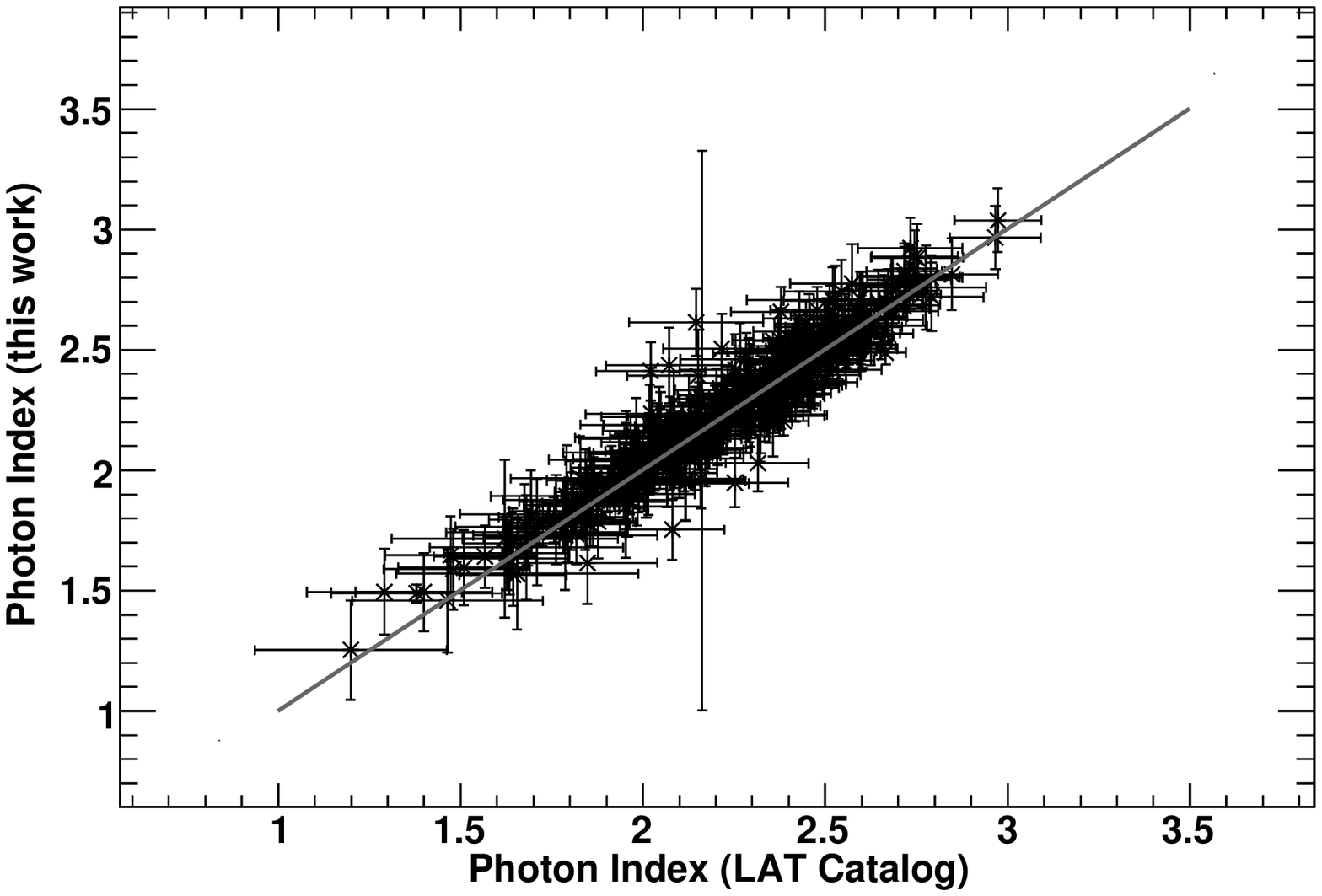}\\
\end{tabular}
  \end{center}
  \caption{Performance of the detection pipeline used
in this work with respect to the detection pipeline used in \cite{cat1}.
The left panel shows the comparison of the reconstructed
$\gamma$-ray fluxes while the right panel shows the comparison
of the photon indices. In both cases the solid line shows the
the locus of points for which the quantity reported in the
y-axis equals the one in the x-axis.
}
  \label{fig:comparison}
\end{figure*}

In order to test the reliability of our detection pipeline
we applied it the to real 1 year dataset. Our aim was to cross
check our result with the result reported in \cite{cat1}.
The flux above 100\,MeV, computed from the power-law
fit to the 100\,MeV--100\,GeV data, is not reported in \cite{cat1}, but
it can be readily obtained using the following expression:
\begin{equation}
F_{100}=E_{piv}\times F_{density}\times \left( \frac{100}{E_{piv}}\right)^{1-\Gamma} \times |1-\Gamma|^{-1},
\end{equation}
where $F_{100}$ is the 100\,MeV--100\,GeV photon flux, $\Gamma$ is 
the absolute value of the photon index,
$E_{piv}$ is the pivot energy and $F_{density}$ is the
flux density at the pivot energy \citep[see][for details]{cat1}.
Fig.~\ref{fig:comparison} shows the comparisons of both
fluxes (above 100\, MeV) and of the photon indices for the
sources detected in both pipelines.  It is clear that 
the fluxes and photon indices derived in this analysis are
reliable; for each source they are consistent with those
in  \cite{cat1} within the reported errors.

The number of sources detected in the simplified pipeline is smaller than
found by \cite{cat1}. Above a TS of 50 and $|b|\geq20^{\circ}$
our approach detects 425 sources while the 1FGL catalog has 
497. Indeed, our aim is not to produce a detection
algorithm which is as sensitive than the one used in \cite{cat1}, but
a detection algorithm which is proven to be 
reliable and can be applied consistently
to both real data and simulations. This allows us to assess properly
all selection effects important for the LAT survey and its analysis.
On this note we remark that all
the 425 sources detected by our pipeline are also detected by \cite{cat1}.
For this reason we limit the studies presented in this work to the
subsample of sources which is detected by our pipeline.
The details of this sample of sources are reported in Tab.~\ref{tab:sample}.
The associations are the ones reported in \cite{agn_cat} and \cite{cat1}.
In our sample 161 sources are classified as FSRQs and 163 as BL Lac objects
while only 4 as blazars of uncertain classification. The number
of sources which are unassociated is 56, thus the identification
incompleteness of this sample is $\sim$13\%.

\begin{deluxetable}{lc}
\tablewidth{0pt}
\tablecaption{Composition of the $|$b$|\geq$20 TS$\geq$50 sample used in this
analysis.
\label{tab:sample}}
\tablehead{
\colhead{CLASS} & \colhead{\# objects }}
\startdata
Total                         & 425 \\
FSRQs                         & 161\\
BL Lacs                       & 163\\
Uncertain\tablenotemark{a}    & 4\\
Blazar Candidates             & 24\\  
Radio Galaxies                & 2 \\
Pulsars                       & 9 \\
Others\tablenotemark{b}       & 6 \\
Unassociated  sources         & 56 \\

\enddata
\tablenotetext{a}{Blazars with uncertain classification.}
\tablenotetext{b}{It includes  Starburst galaxies, 
Narrow line Seyfert 1 objects and Seyfert galaxy  candidates.}
\end{deluxetable}

\subsection{Derivation of the Sky Coverage}
\label{sec:skycov}

\begin{figure}[h!]
\begin{centering}
	\includegraphics[scale=0.6]{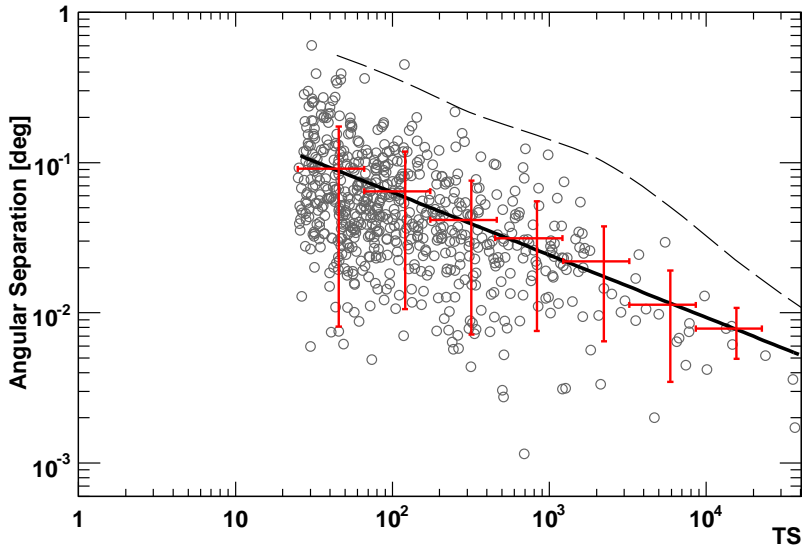} 
	\caption{Angular separation of the real LAT sources from the most
probable associated counterpart as a function of TS. All sources
with $|b|\geq$10$^{\circ}$
with a probability of associations larger than 0.5 were used 
\citep[see ][for a definition of probability of association]{agn_cat}. The
solid line is the best fit for the mean offset of the angular separations
while the dashed line represents the observed  5\,$\sigma$ error radius
as a function of test statistics. Note that the 5\,$\sigma$ error radius
is weakly dependent on the level of probability of association chosen.}
	\label{fig:angsep}
\end{centering}
\end{figure}

In order to derive the sky coverage from simulations,
detected sources (output) need to be associated to the simulated
ones (input). We do this on a statistical basis using an estimator
which is defined for each set of input-output sources as:
\begin{equation}
R^2 = \left( \frac{||\bar{x}-\bar{x_0}||}{\sigma_{pos}} \right)^2 + 
\left( \frac{S-S_0}{\sigma_S} \right)^2 + 
\left( \frac{\Gamma-\Gamma_0}{\sigma_{\Gamma}}  \right)^2
\end{equation}
where $\bar{x}$, $S$ and $\Gamma$ are the source coordinates, fluxes 
and photon indices as determined from the ML step while
$\bar{x_0}$, $S_0$ and $\Gamma_0$ are the simulated (input) values.
The 1\,$\sigma$ errors on the position, flux and photon index
are $\sigma_{pos}$, $\sigma_S$ and $\sigma_{\Gamma}$ respectively.
We then  flagged as the  most likely associations
those pairs with the minimum value of R$^2$.
All pairs with an angular separation which is larger than the 4\,$\sigma$
error radius
are flagged as spurious and excised from
the following analysis. The empirical, as derived from the real
data, 5\,$\sigma$ error radius as a function of source TS is shown in
Fig.~\ref{fig:angsep}.
 As in \cite{hasinger93} and in \cite{cappelluti07}
we defined {\it confused} sources for which the ratio  
$S/(S_0+3\sigma_S)$ (where $\sigma_S$ is the error on the output flux)
is larger than 1.5.  We found  that,
according to this criterion, $\sim$4\,\% of the 
sources (detected for $|b|\geq10^{\circ}$) are confused in the first year survey.

\begin{figure*}[ht!]
  \begin{center}
  \begin{tabular}{cc}
    \includegraphics[scale=0.4]{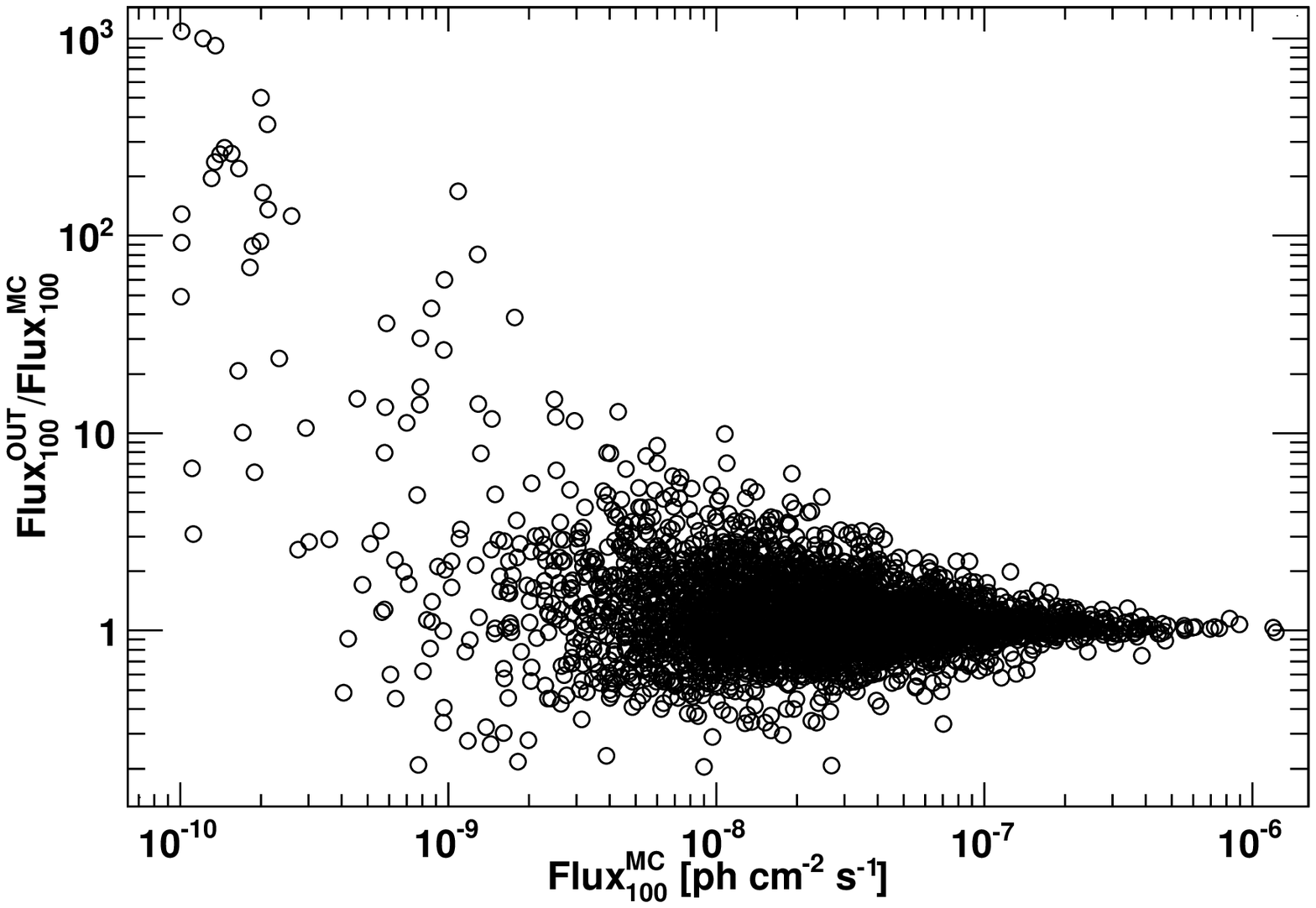} 
	 \includegraphics[scale=0.4]{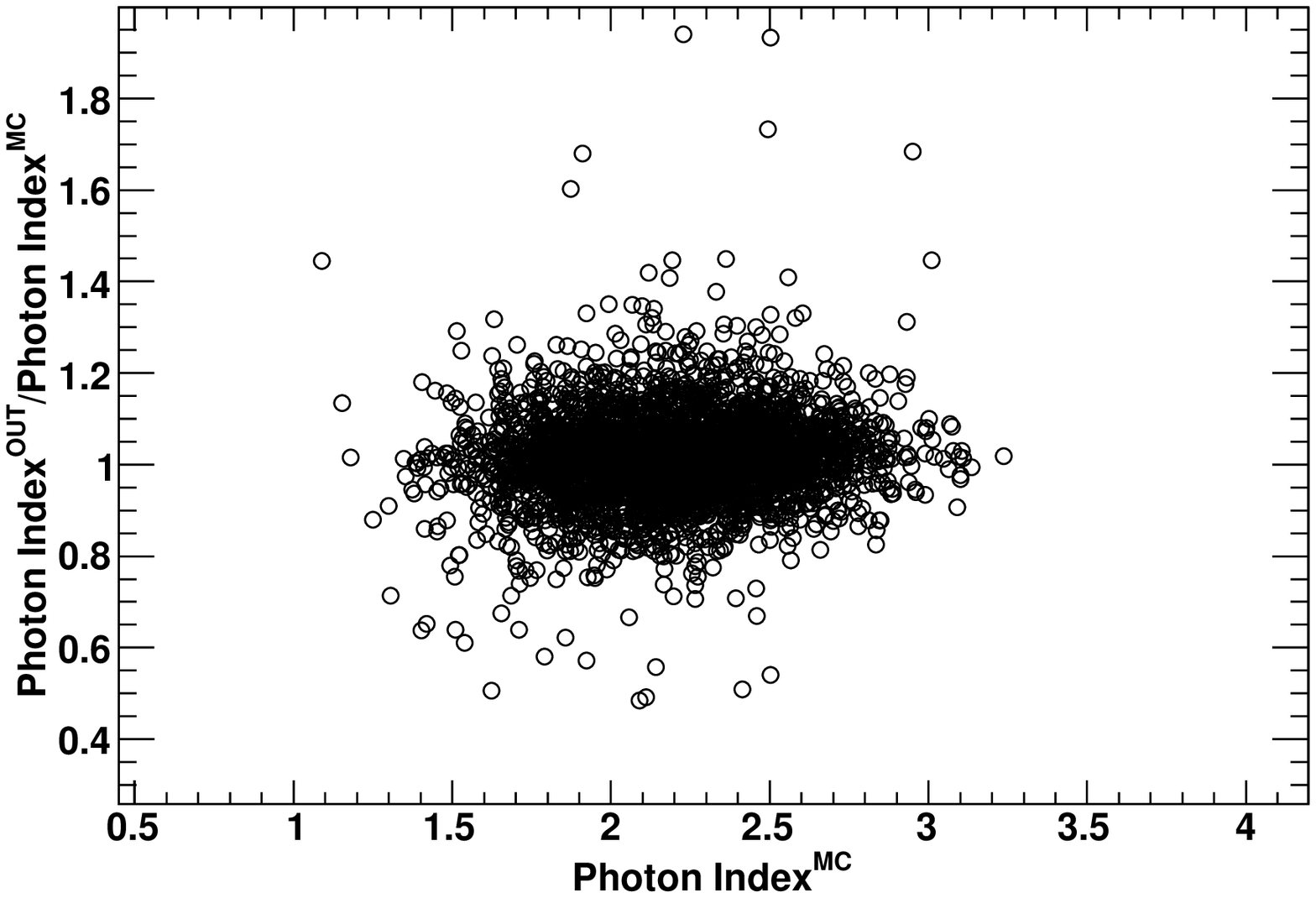}\\
\end{tabular}
  \end{center}
  \caption{Left Panel: Reconstructed versus Simulated fluxes
for all sources with TS$\geq$50 and $|b$$|\geq$20$^{\circ}$.
For the analysis reported here only sources with 
F$_{100}\geq10^{-9}$ ph cm$^{-2}$ s$^{-1}$ are considered.
Right Panel: Reconstructed versus Simulated photon indices for 
all sources with TS$\geq$50 and $|b$$|\geq$20$^{\circ}$.
}
  \label{fig:simulations}
\end{figure*}

The right panel of Fig.~\ref{fig:simulations}  shows
the ratio of reconstructed to simulated source flux versus
 the  simulated source flux.
At medium to bright fluxes the distribution of the ratio is  centered on unity
showing that there are no systematic errors in the flux measurement.
At low fluxes (in particular for F$_{100}<10^{-9}$\,ph cm$^{-2}$ s$^{-1}$)
the distribution is
  slightly (or somewhat) biased toward values greater than unity.
This is  produced by three effects:
1) source confusion, 2) Eddington bias \citep{eddington40} and
3) non converging Maximum Likelihood fits (see $\S$~\ref{sec:mlfit} 
for details).
The Eddington bias arises from  measurement errors of any
intrinsic source property (e.g. source flux). Given its nature,
it affects only sources close to the detection threshold.
Indeed, at the detection threshold the uncertainty in the reconstructed
fluxes makes sources with a measured flux slightly larger than
the real value more easily detectable in the survey rather
than those with a measured flux slightly lower than the real one.
This causes the shift of the flux ratio
 distribution of Fig.~\ref{fig:simulations} to move systematically
 to values larger than unity at low fluxes.
In any case, the effect of this bias is not relevant as it affects
less than 1\,\% of the entire population. 
This uncertainty will be neglected as only sources 
with F$_{100}\geq10^{-9}$\,ph cm$^{-2}$ s$^{-1}$ will be considered for the 
analysis presented here.
Moreover, the right panel of Fig.~\ref{fig:simulations} shows that 
the measured photon index agrees well with the simulated one.

\begin{figure}[h!]
\begin{centering}
	\includegraphics[scale=0.6]{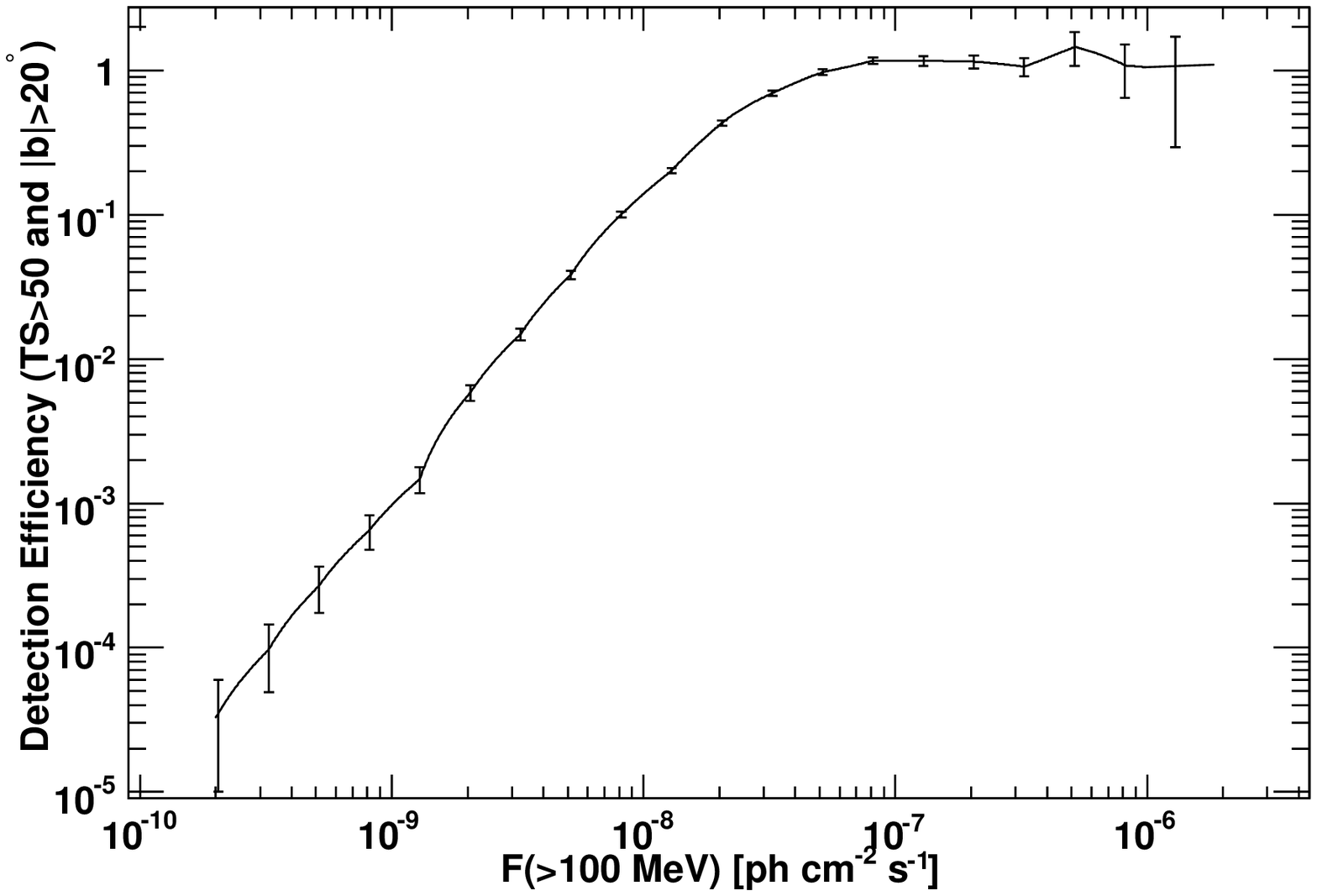} 
	\caption{Detection efficiency as a function of measured source flux for
$|b|\geq20^{\circ}$,  TS$\geq50$ and for a sample of sources
with a mean photon index of 2.40 and dispersion of 0.28. The
error bars represent statistical uncertainties from the 
counting statistic of our Monte Carlo simulations.}
	\label{fig:skycov}
\end{centering}
\end{figure}

In addition to assessing the reliability and biases of our source
detection procedure, the main aim of these simulations is to provide
a precise estimate of the completeness function of the {\it Fermi}/LAT
survey (known also as sky coverage). The one-dimensional 
sky coverage can be derived for each bin of flux as the ratio
between the number of detected sources and the number of simulated sources.
The detection efficiency for the entire TS$\geq50$ and $|b|\geq20^{\circ}$
sample is reported in Fig.~\ref{fig:skycov}.
This plot shows that the LAT sensitivity
extends all the way to F$_{100}\sim10^{-10}$\,ph cm$^{-2}$ s$^{-1}$
although at those fluxes only the hardest sources can be detected.
Also the sample becomes complete for 
 F$_{100}=7-8\times 10^{-8}$\,ph cm$^{-2}$ s$^{-1}$.
Since for these simulations, the {\it intrinsic} distribution of 
photon indices has been used (see e.g. $\S$~\ref{sec:photon})
this sky coverage takes properly into  account the bias towards
the detection of hard sources. This also means that this
sky coverage cannot be applied to other source samples with very
different photon index distributions.

%
%
\section{Systematic Uncertainties}
\label{sec:syst}

\subsection{Non converging Maximum Likelihood fits}
\label{sec:mlfit}
A small number of sources detected by our pipeline have
unreliable spectral fits. Most of the time, these sources
have a reconstructed photon index which is very soft (e.g. $\sim$5.0)
and at the limit of the accepted range of values. As a consequence
their reconstructed flux overestimates the true flux by up to factor 1000
(see left panel of Fig.~\ref{fig:simulations}).
This is due to the fact
the the ML algorithm does not find an absolute minimum of the fitting
function for these cases. 
Inspection of the
regions of interests (ROIs) of these objects shows that this tends
to happen either in regions very dense with sources
or  close to the Galactic plane, where the diffuse emission is the brightest.
The best approach in this case would be to adopt an iterative procedure
for deriving the best-fitting parameters which starts by optimizing
the most intense components (e.g. diffuse emissions and bright sources)
and then move to the fainter ones. This procedure is correctly implemented
in \cite{cat1}. Its application to our problem would make the processing
time of our simulations very long and we  note that
the systematic uncertainty deriving from it is small.
Indeed, the number of sources with unreliable spectral parameters
are for $TS\geq25$ are 2.3\,\% and 2.0\,\%  for $|b|\geq15^{\circ}$ a
$|b|\geq20^{\circ}$ respectively. 
These fractions decrease to 1.2\,\% and 0.9\,\% adopting TS$\geq50$.

To limit the systematic uncertainties in this analysis,
we will thus  select only those sources
which are detected above TS$\geq50$ and $|b|\geq20^{\circ}$.
It will also be shown that results do not change
if the sample is enlarged to include all sources with $|b|\geq15^{\circ}$.

\subsection{Variability}
It is well known that blazars are inherently variable objects with variability
in flux of up to a factor 10 or more. Throughout  this work
only average quantities (i.e. mean flux and mean photon index) are used.
This is correct in the context of the determination of the mean energy
release in the Universe of each source. Adopting the peak flux (i.e. 
the brightest flux displayed by each single source) would produce the net
effect of overestimating the true intrinsic source density at any flux 
\citep[see the examples in][]{reimer01} with the result of overestimating
the contribution of sources to the diffuse background. 

It is not straightforward to determine how blazar variability affects
the analysis presented here. 
On timescales large enough (such as the one
spanned by this analysis), the mean flux is a good estimator of
the mean energy release of a source. This is not true anymore on
short timescales (e.g. $\sim$1\,month) 
since the mean flux corresponds to the source
flux at the moment of the observation. The continuous scanning
of the $\gamma$-ray sky performed by
 {\it Fermi} allows to determine long-term variability
with unprecedented accuracy. As shown already in \cite{lat_lbas} the 
picture arising from {\it Fermi} is rather different from the one derived by
EGRET \citep{hartman99}. Indeed, the peak-to-mean flux ratio 
for {\it Fermi} sources is considerably smaller than for EGRET sources. 
For most
of the {\it Fermi} sources this is just a factor 2, as is confirmed
in the 1\,year sample \citep[see Fig.10 in][]{agn_cat}. This excludes the 
possibility that most of the sources are detected because of a single
outburst which happened during the 11\,months of observation and
are undetected for the remaining time.  Moreover, as shown in 
\cite{sed} there is little or no variation of the photon index
with flux. We thus believe that no large systematic uncertainties
are derived from the use of average physical quantities and the
total systematic uncertainty (see next section) will be slightly overestimated
to accommodate possible uncertainties caused by variability.

\subsection{Non power law spectra}
\label{sec:pow}
\begin{figure}[h!]
\begin{centering}
	\includegraphics[scale=0.6]{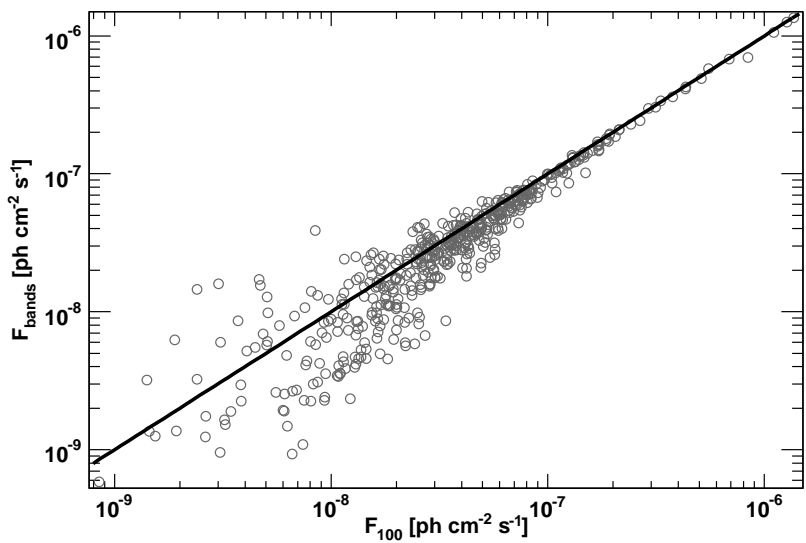} 
	\caption{Source flux estimated with a power-law fit to the 0.1--100\,GeV
band versus the sum of the source fluxes derived in 5 contiguous energy bands 
\cite[see][for details]{cat1}. The solid line is the F$_{bands}$=F$_{100}$
relation. The spread at low fluxes arises from the difficulties of 
estimating the source flux in small energy bands.
}
	\label{fig:fluxcomp}
\end{centering}
\end{figure}

It is well known that the spectra of blazars are complex and often show
curvature when analyzed over a large waveband. In this case the
approximation of their spectrum with a simple power law (in the
0.1--100\,GeV band) might provide a poor estimate  of their
true flux. To estimate the uncertainties derived by this assumption
we plotted for the extragalactic sample used here (e.g. TS$\geq$ 50
and $|b|\geq$20$^{\circ}$) the source flux as derived
from the power-law fit to the whole band versus  the source flux 
as derived from the sum of the fluxes  over the 5 energy bands
reported in \cite{cat1}.  This comparison is reported 
in Fig.~\ref{fig:fluxcomp}. From the figure it is apparent that
the flux (F$_{100}$) derived from a power-law fit to the
whole band overestimates slightly the true source flux.
Analysis of the ratio between the power-law flux and flux derived in
5 energy bands, shows that on average the F$_{100}$ flux overestimates
the true source flux by $\sim$8\,\%. At very bright fluxes (e.g. 
F$_{100}\geq10^{-7}$\,ph cm$^{-2}$ s$^{-1}$) the overestimate reduces
to $~\sim$5\,\%. For the analysis presented here we will thus assume
that the total systematic uncertainty connected to the use of fluxes
computed  with a power-law fit over the broad 0.1--100\,GeV band is 8\,\%.

Considering also the uncertainties of the previous sections,
we derive that the total systematic uncertainty for the sample
used here (TS$\geq$50 and $|b|\geq$20$^{\circ}$) is $\sim$10\,\%.
Since this uncertainty affects mostly the determination of the source flux
it will be propagated by shifting in flux
 the sky coverage of Fig.\ref{fig:skycov}
by $\pm10$\,\%.

\section{Source Counts Distributions}
\label{sec:logn}

The source counts distribution, commonly referred to as log $N$--log$S$
or size distribution, is the cumulative number of sources $N(>S)$
detected above a given flux $S$. In this section we apply
several methods to derive the source count distribution of 
{\it Fermi}/LAT sources. We also remark that the catalog used
for this analysis is the one described in $\S$~\ref{sec:cat} (see also
Tab.~\ref{tab:sample}).

\subsection{Standard Approach}
\label{sec:logn_1d}

A standard way to derive the (differential) log $N$--log $S$ is 
through the following expression:
\begin{equation}
\frac{dN}{dS} = \frac{1}{\Delta\ S}\
\sum_{i=1}^{N_{\Delta S}} \frac{1}{\Omega_i}
\end{equation}

where $N_{\Delta S}$ is the total number of detected sources with fluxes
in the $\Delta$S interval, and $\Omega_i$ 
 is the solid angle associated
with the flux of the $i_{th}$ source (i.e.,
 the detection efficiency multiplied by the survey solid angle).
We also note that formally $N$ is an areal density and should
be expressed as $dN/d\Omega$. However for simplicity of notation
the areal density will, throughout this paper, be expressed as $N$.
For the $|b|\geq20^{\circ}$ sample  the geometric solid angle
of the survey is 27143.6\,deg$^{2}$.
In each flux bin, the final uncertainty is obtained by summing in quadrature
the error on the number of sources and the systematic uncertainties
described in $\S$~\ref{sec:syst}.

\begin{figure*}[ht!]
  \begin{center}
  \begin{tabular}{cc}
    \includegraphics[scale=0.4]{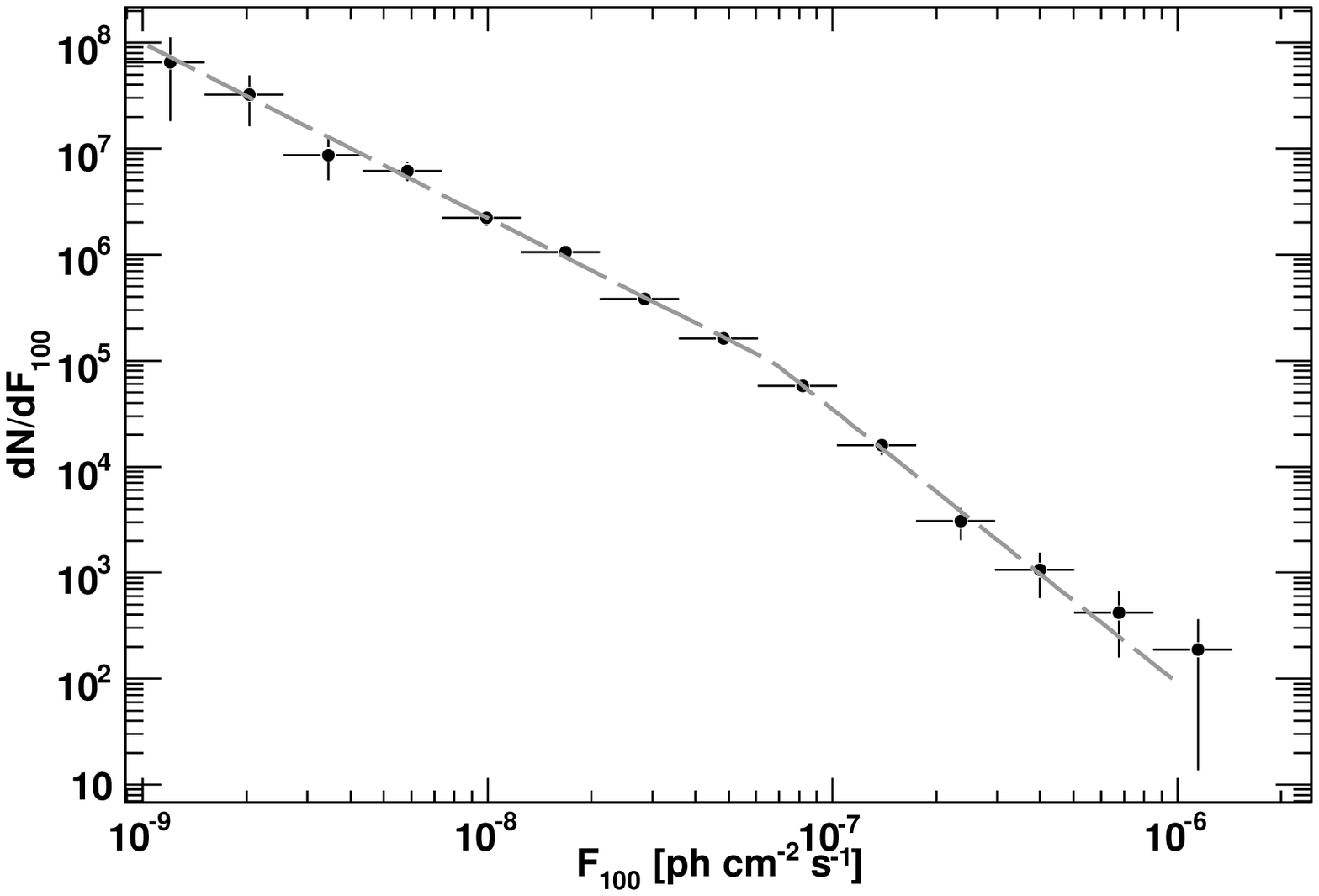} 
	 \includegraphics[scale=0.4]{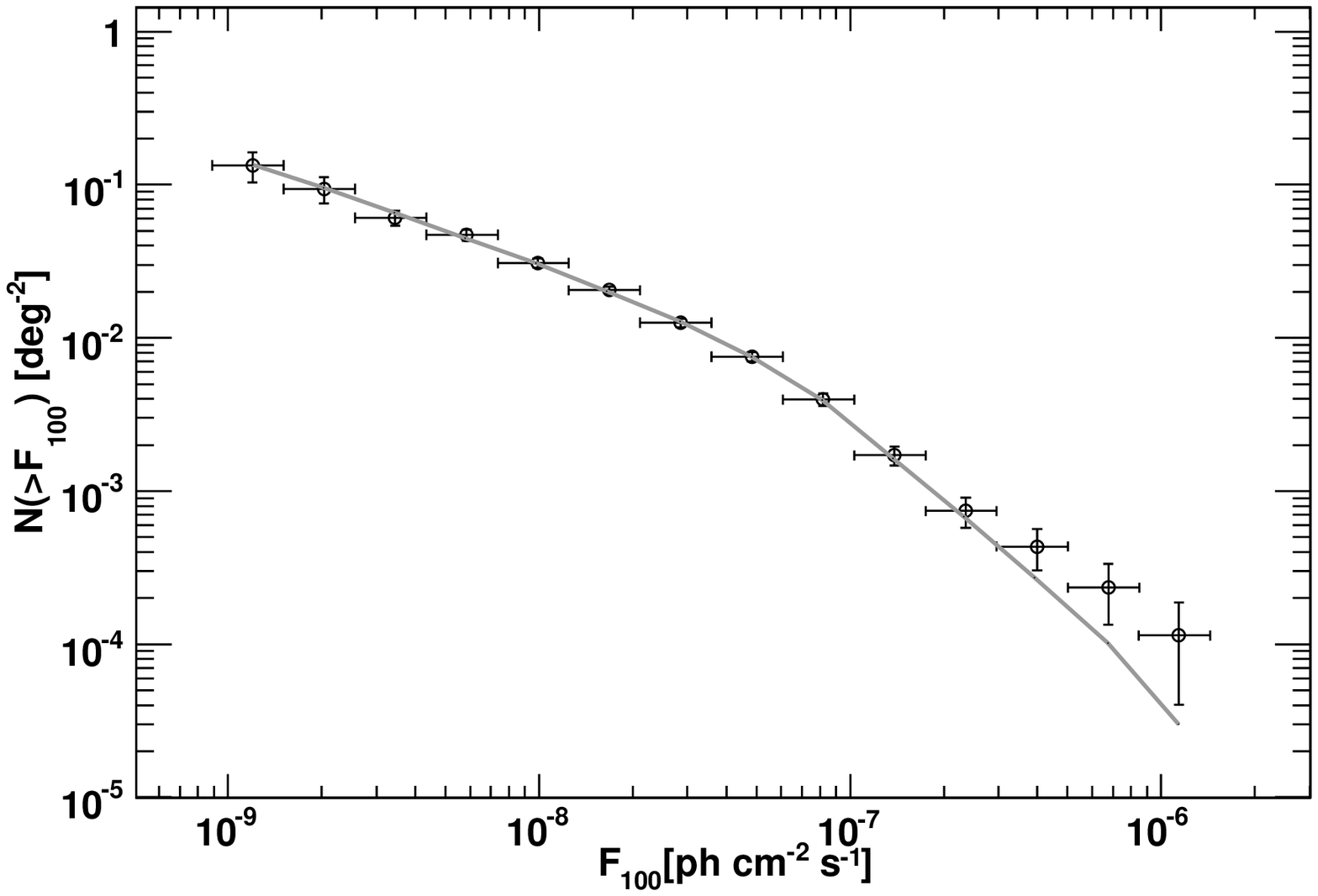}\\
\end{tabular}
  \end{center}
  \caption{Differential (left) and cumulative (right)
log $N$--log $S$ for all sources with TS$\geq50$ and $|b|\geq20^{\circ}$.
The dashed line is the best-fit broken power law model as reported in
the text.}
  \label{fig:logn_1d}
\end{figure*}

Both the differential and the cumulative version of the source
count distributions are reported in Fig.~\ref{fig:logn_1d}.
In order to parametrize the source count distribution
we perform a $\chi^{2}$ fit to the differential data using 
a broken power-law model of the type:
\begin{eqnarray}
\label{eq:dblpow}
\frac{dN}{dS} & = & A S^{-\beta_1} \ \ \ \ \ \ \ \ \ \ S \geq S_b \nonumber \\
       & = & A S_b^{-\beta_1+\beta_2}S^{-\beta_2} \ \ S < S_b 
\end{eqnarray}

where $A$ is the normalization and $S_b$ is the flux break.
The best-fit parameters are reported in Tab.~\ref{tab:logn_1d}.
The log $N$-- log $S$ distribution
of GeV sources shows a strong break ($\Delta \beta=\beta_1 -\beta_2\approx1.0$)
at F$_{100} =6.97(\pm0.13)\times10^{-8}$\,ph cm$^{-2}$ s$^{-1}$. At fluxes
brighter than the break flux, the source count distribution is 
consistent with Euclidean ($\beta_1 = 2.5$)
while it is not at fainter fluxes.
As Tab.~\ref{tab:logn_1d} shows, 
these results do not change if 
the sample under study is enlarged to $|b|\geq$15$^{\circ}$.

\begin{deluxetable}{lcccccccc}
\tablewidth{0pt}
\tabletypesize{\footnotesize}
\tablecaption{
Results of the power-law fits to the differential source count
distributions obtained with the standard  method of $\S$~\ref{sec:logn_1d}
\label{tab:logn_1d}}
\tablehead{
\colhead{} & \colhead{} & \multicolumn{2}{c}{Sample Limits} & \colhead{} &
\multicolumn{4}{c}{Best-fit Parameters}\\
\cline{3-4} \cline{6-9} \\ 
\colhead{SAMPLE} & \colhead{\# Objects } &
\colhead{TS$\geq$}     & \colhead{$|$b$|\geq$} & \colhead{} &
\colhead{A\tablenotemark{a}} &  \colhead{$\beta_1$} &
\colhead{S$_b$\tablenotemark{b}}  & \colhead{$\beta_2$} 
}
\startdata
ALL & 425 & 50 & 20$^{\circ}$ & & 1.15$^{+0.15}_{-0.15}$ & 2.63$^{+0.22}_{-0.19}$ & 6.97$^{+1.28}_{-1.29}$ & 1.64$^{+0.06}_{-0.07}$\\
ALL & 483 & 50 &  15$^{\circ}$ & & 1.74$^{+0.16}_{-0.16}$ & 2.60$^{+0.19}_{-0.17}$ & 6.40$^{+1.04}_{-1.08}$ & 1.60$^{+0.06}_{-0.07}$

\enddata
\tablenotetext{a}{In units of 10$^{-14}$\,cm$^{2}$ s deg$^{-2}$.}
\tablenotetext{b}{In units of $10^{-8}$\,ph cm$^{-2}$ s$^{-1}$ 
(0.1$\leq$E$\leq$100\,GeV).}

\end{deluxetable}

\subsection{A Global Fit}
\label{sec:logn_2d}

Because of the spectral selection effect discussed in $\S$~\ref{sec:photon},
the sky coverage derived in $\S$~\ref{sec:skycov} can  be used only
with samples which have a distribution of the photon indices similar
to the one used in the simulations (i.e. a Gaussian with mean and dispersion
of 2.40 and 0.28). Here we aim at overcoming this limitation by implementing
for the first time a novel,
more formal, analysis to derive the source count distribution.
We aim at describing the  properties of the sample in terms of
a distribution function of the following kind:
\begin{equation}
\frac{dN}{dSd\Gamma} = f(S) \cdot g(\Gamma)
\label{eq:dn2}
\end{equation}

where  $f(S)$ is  the intrinsic flux distribution of sources
and $g(\Gamma)$ is the intrinsic distribution of the 
photon indices.
In this analysis, $f(S)$ is modeled as a double power-law function
as in Eq.~\ref{eq:dblpow}.
The index distribution $g(\Gamma)$ is modeled a Gaussian function:
\begin{equation}
g(\Gamma) = e^{-\frac{ (\Gamma-\mu)^2}{2\sigma^2}}
\end{equation}

where $\mu$ and $\sigma$ are respectively the mean and the dispersion
of the Gaussian distribution. As it is clear from Eq.~\ref{eq:dn2},
we made the hypothesis that the $dN/dSd\Gamma$ function
is factorizable in two separate distributions in flux and photon index.
This is the most simple assumption that could be made and as it will be
shown in the next sections it provides a good description of the data.
Moreover, we emphasize, as already did in $\S$~\ref{sec:sim}, that
this analysis implicitly assumes that the photon index distribution
does not change with flux. This will be discussed in more details
in the next sections.

This function is then fitted to all datapoints using a Maximum Likelihood
approach as described in Sec.~3.2 of \cite{ajello09b}.
In this method, the Likelihood function can be defined as:
\begin{equation}
L = {\rm exp(-N_{exp})} \prod_{i=1}^{N_{\rm obs}}\lambda (S_i,\Gamma_i)
\end{equation}

with $\lambda (S,\Gamma)$ defined as:

\begin{equation}
\lambda (S,\Gamma) = \frac{dN}{dSd\Gamma}\Omega(S,\Gamma)
\end{equation}

where $\Omega(S,\Gamma)$ is the photon index dependent sky coverage
and $N_{\rm obs}$ is the number of observed sources.
This is generated from the same Monte Carlo simulation of $\S$~\ref{sec:sim}
with the difference that this time the detection probability is computed
for each bin of the photon-index--flux plane as the ratio between 
detected and simulated sources (in that bin). This produces a
sky coverage which is function of both the source flux and photon index.

The {\it expected} number of sources $N_{exp}$ can be computed
as:
\begin{equation}
N_{exp}=\int d\Gamma \int dS \lambda (S,\Gamma)
\end{equation}

The maximum likelihood parameters are obtained by minimizing the function
$C(=-2 {\rm ln} L)$:
\begin{equation}
C = -2 \sum^{N_{obs}}_i {\rm ln} (\lambda(S_i,\Gamma_i)) 
- 2N\ {\rm ln(N_{exp} ) }
\end{equation}

while  their associated 1\,$\sigma$ errors are computed by varying
the parameter of interest, while the others are allowed to float,
until an increment of $\Delta C$=1 is achieved. This gives
an estimate of the 68\,\% confidence region for the parameter of interest
\citep{avni76}.

Once the $dN/dSd\Gamma$ has been determined, the standard differential
source count distribution can be readily derived as:
\begin{equation}
\frac{dN}{dS} = \int_{-\infty}^{\infty} d\Gamma \frac{dN}{dS d\Gamma}
\end{equation}

\subsection{The Total Sample of Point Sources}

\begin{figure*}[ht!]
  \begin{center}
  \begin{tabular}{cc}
    \includegraphics[scale=0.4]{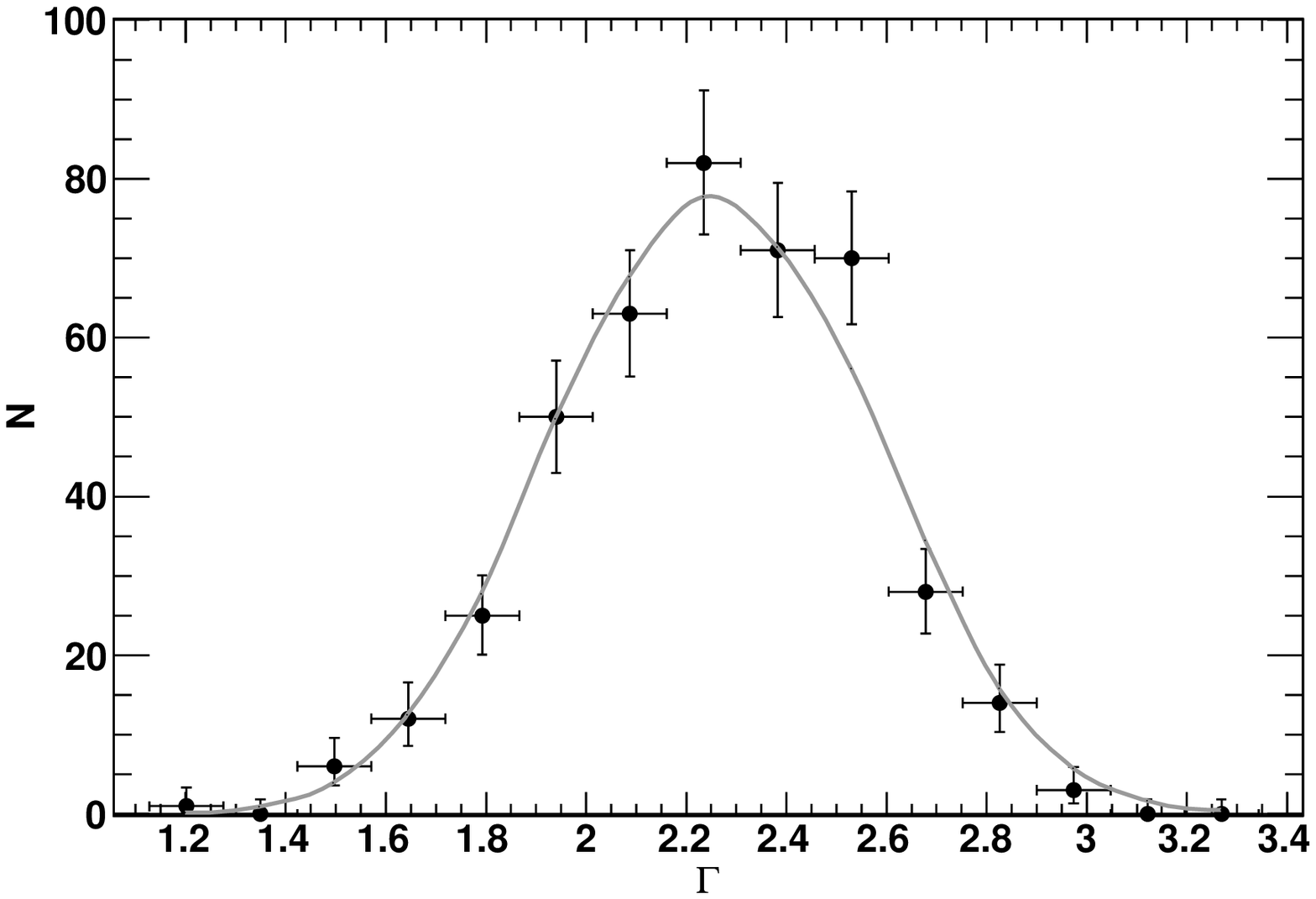} 
	 \includegraphics[scale=0.4]{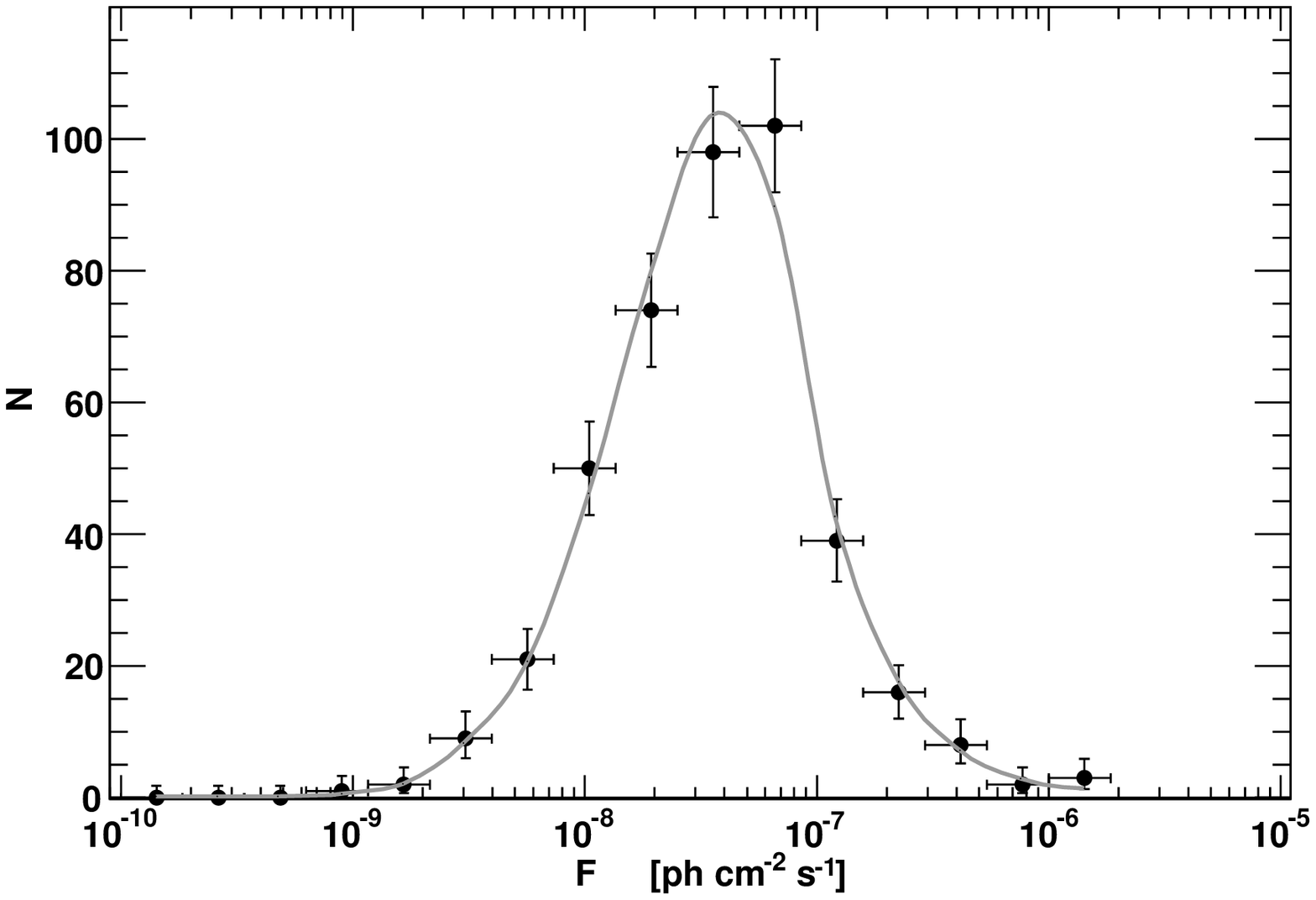}\\
\end{tabular}
  \end{center}
  \caption{Distribution of photon indices (left) and fluxes (right)
for the TS$\geq$50 and $|b|\geq$20$^{\circ}$ sources. The dashed
line is the best fit $dN/dSd\Gamma$ model. Using the $\chi^{2}$ test
the probabilities that the data and the model line come from the
same parent population are 0.98 and 0.97 for the photon index 
and flux distribution respectively.}
  \label{fig:total_distr}
\end{figure*}

\begin{figure}[h!]
\begin{centering}
	\includegraphics[scale=0.6]{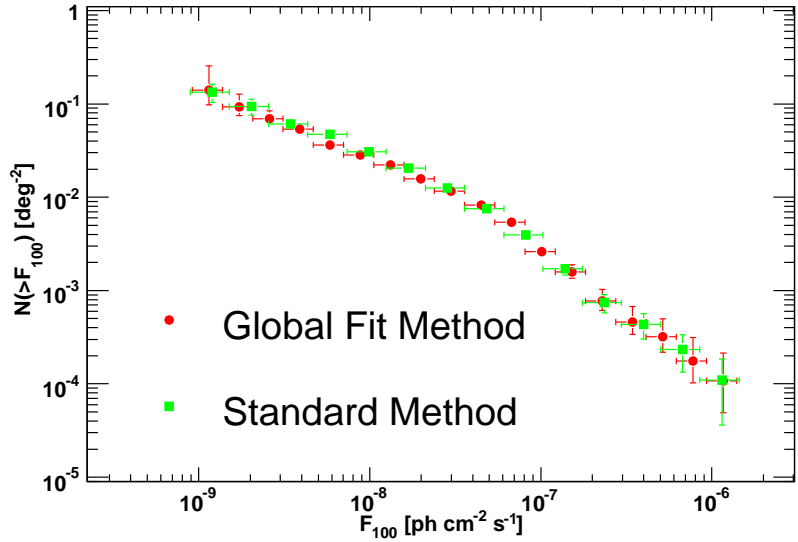} 
	\caption{Comparison of log $N$--log $S$ of the whole sample of
(TS$\geq$50 and $|b|\geq$20$^{\circ}$) sources
built with the
standard method (green datapoints, see $\S$~\ref{sec:logn_1d}) 
and the global fit method (red datapoints,see $\S$~\ref{sec:logn_2d}).
}
	\label{fig:lognboth}
\end{centering}
\end{figure}

The results of the best-fit model for the entire sample of sources
(for TS$\geq$50 and $|b|\geq$20$^{\circ}$) are reported in
Tab.~\ref{tab:logn2D}.
Fig.~\ref{fig:total_distr} shows how well the best-fit model
reproduces the observed index and flux distributions.
The $\chi^2$ test yields that the probabilities that the 
real distribution and the model line come from the same
parent population are 0.98 and 0.97 for the photon index and flux 
distributions, respectively. 
In Fig.~\ref{fig:lognboth} the source count distribution obtained
here is compared to the one derived using the standard approach
of $\S$~\ref{sec:logn_1d}; the good agreement is apparent.

\begin{figure}[h!]
\begin{centering}
	\includegraphics[scale=0.6]{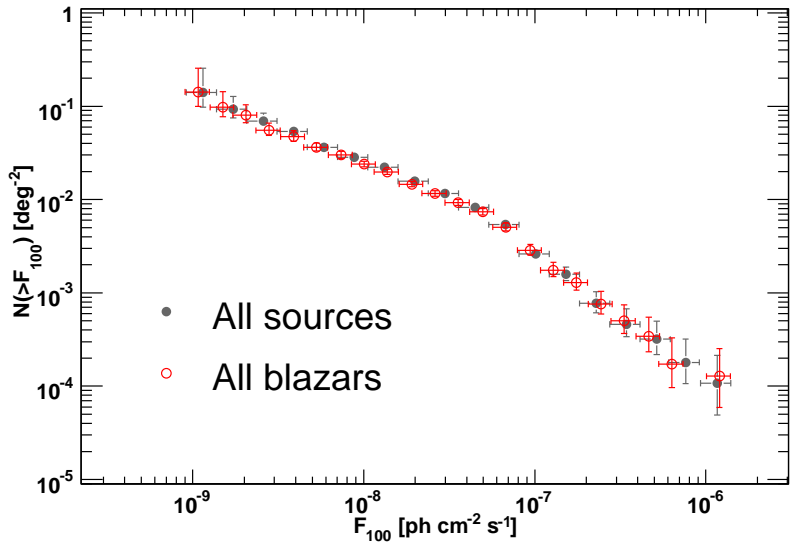} 
	\caption{Comparison between log $N$--log $S$ distributions
 of the whole sample of sources (solid circles) and blazars (open circles).
The solid line are the respective best-fit models as reported in
Tab.~\ref{tab:logn2D}.
}
	\label{fig:lognblaz}
\end{centering}
\end{figure}

We also derived the source count distributions of all objects
which are classified as blazars (or candidate blazars)
in our sample. This includes 352 out of the 425 objects reported
in Tab.~\ref{tab:sample}. The number of sources
that lack association is 56 and thus the incompleteness
of the blazar sample is 56/425$\approx 13$\,\%. A reasonable and simple
assumption is that the 56 unassociated sources are distributed among the
different source classes in a similar way as the associated portion 
of the sample (see Tab.~\ref{tab:sample}).
This means that $~$46 out of the 56 unassociated sources  are likely
to be blazars.
As it is possible to notice both from the best-fit
parameters of Tab.~\ref{tab:logn2D} and from Fig.~\ref{fig:lognblaz},
there is very little difference between the source count distributions
of the entire sample and the one of blazars. This confirms on a 
statistical basis that
most of the 56 sources without association are  likely to be blazars.
It is also clear from Fig.~\ref{fig:total_distr},
that the model (e.g. Eq.~\ref{eq:dn2})
represents a satisfactory description of the data.
This also implies that the {\it intrinsic} photon index distribution of blazars
is compatible with a Gaussian distribution that does not change
(at least dramatically) with source flux in the range of fluxes spanned 
by this analysis.
A change in the average spectral properties of blazars with flux 
(and/or redshift) might be caused
by the different cosmological evolutions of FSRQs and BL Lacs
or by the spectral evolution of the two source classes with redshift.
While it is something reasonable to expect, this effect is in the current
dataset not observed. The luminosity function, which is left to
a future paper, will allow us to investigate this effect 
in great detail.

\begin{deluxetable}{lccccccccccc}
\tablewidth{0pt}
\rotate
\tabletypesize{\scriptsize}
\tablecaption{
Results of the best fits to the source count
distributions.
\label{tab:logn2D}}
\tablehead{
\colhead{} & \colhead{} & \colhead{} &
 \multicolumn{2}{c}{Sample Limits} & \colhead{} &
\multicolumn{6}{c}{Best-fit Parameters}\\
\cline{4-5} \cline{7-12} \\ 
\colhead{SAMPLE} & \colhead{\# Objects } & \colhead{Incompl.} & 
\colhead{TS$\geq$}     & \colhead{$|$b$|\geq$} & \colhead{} &
\colhead{A\tablenotemark{a}} &  \colhead{$\beta_1$} &
\colhead{S$_b$\tablenotemark{b}}  & \colhead{$\beta_2$} &
\colhead{$\mu$}   &  \colhead{$\sigma$}
}

\startdata
ALL & 425 & 0 & 50 & 20$^{\circ}$ & & 16.46$\pm0.80$ & 2.49$\pm0.12$ & 6.60$\pm0.91$ & 1.58$\pm0.08$ & 2.36$\pm0.02$ & 0.27$\pm0.01$\\

BLAZAR & 352 & 0.13 & 50 & 20$^{\circ}$ & & 18.28$\pm1.00$& 2.48$\pm0.13$ & 7.39$\pm1.01$ & 1.57$\pm0.09$ & 2.37$\pm0.02$ & 0.28$\pm0.01$ \\

FSRQ    & 161 & 0.19 & 50 & 20$^{\circ}$ & &  72.41$\pm5.76$ & 2.41$\pm0.16$ & 6.12$\pm1.30$ & 0.70$\pm0.30$ & 2.48$\pm0.02$ & 0.18$\pm0.01$\\
BL Lac  & 163  & 0.19 & 50 & 20$^{\circ}$ & & 0.106$\pm0.009$ & 2.74$\pm0.30$ & 6.77$\pm1.30$& 1.72$\pm0.14$ & 2.18$\pm0.02$ & 0.23$\pm0.01$ \\
Unassociated   & 56  & 0 & 50 & 20$^{\circ}$ & & 3.12$(\pm0.5)\times10^{-5}$ & 3.16$\pm0.50$ & 4.48$\pm1.3$ & 1.63$\pm0.24$ & 2.29$\pm0.03$ & 0.20

%
%
%
%
%
\enddata
\tablenotetext{a}{In units of 10$^{-14}$\,cm$^{2}$ s deg$^{-2}$.}
\tablenotetext{b}{In units of 10$^{-8}$\,ph cm$^{-2}$ s$^{-1}$.}
%
%
%
\end{deluxetable}

\subsection{FSRQs}
\label{sec:fsrq}
For the classification of blazars as flat spectrum 
radio quasars (FSRQs) or BL Lacertae objects (BL Lacs)
we use the same criteria adopted in \cite{lat_lbas}.
This classification relies on the conventional definition of BL Lac objects outlined in \cite{stocke91}, \cite{urry95}, and \cite{marcha96}
 in which the equivalent width of the strongest optical emission line is 
$<$5\,\AA\, and the optical spectrum shows a Ca II H/K break ratio C$<$0.4. 

It is important to determine correctly the incompleteness
of the sample when dealing with a sub-class of objects.
Indeed, in the sample of Tab.\ref{tab:sample}, 56 objects
have no associations and 28 have either an uncertain or 
a tentative association with blazars. Thus the total incompleteness is
84/425 = $\sim$19\,\% when we refer to either FSRQs or BL Lac objects
separately. The incompleteness levels of all the samples used here 
are for clarity
reported also in Tab.~\ref{tab:logn2D}.
 Since we did not perform dedicated simulations for
the FSRQ and the BL Lac classes, their source count distributions can 
be derived only with the method described in $\S$~\ref{sec:logn_2d}.

\begin{figure*}[ht!]
  \begin{center}
  \begin{tabular}{cc}
    \includegraphics[scale=0.4]{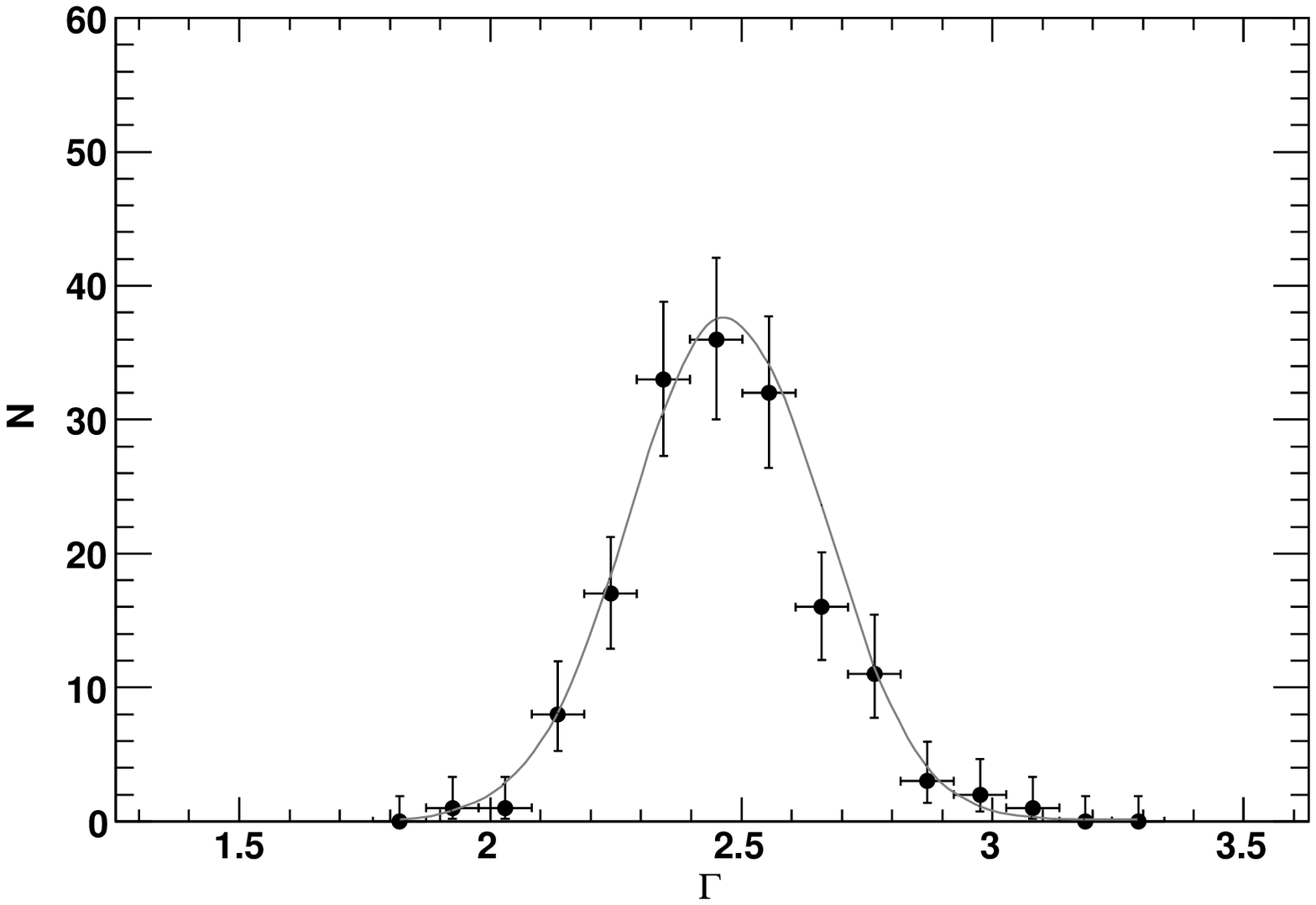} 
	 \includegraphics[scale=0.4]{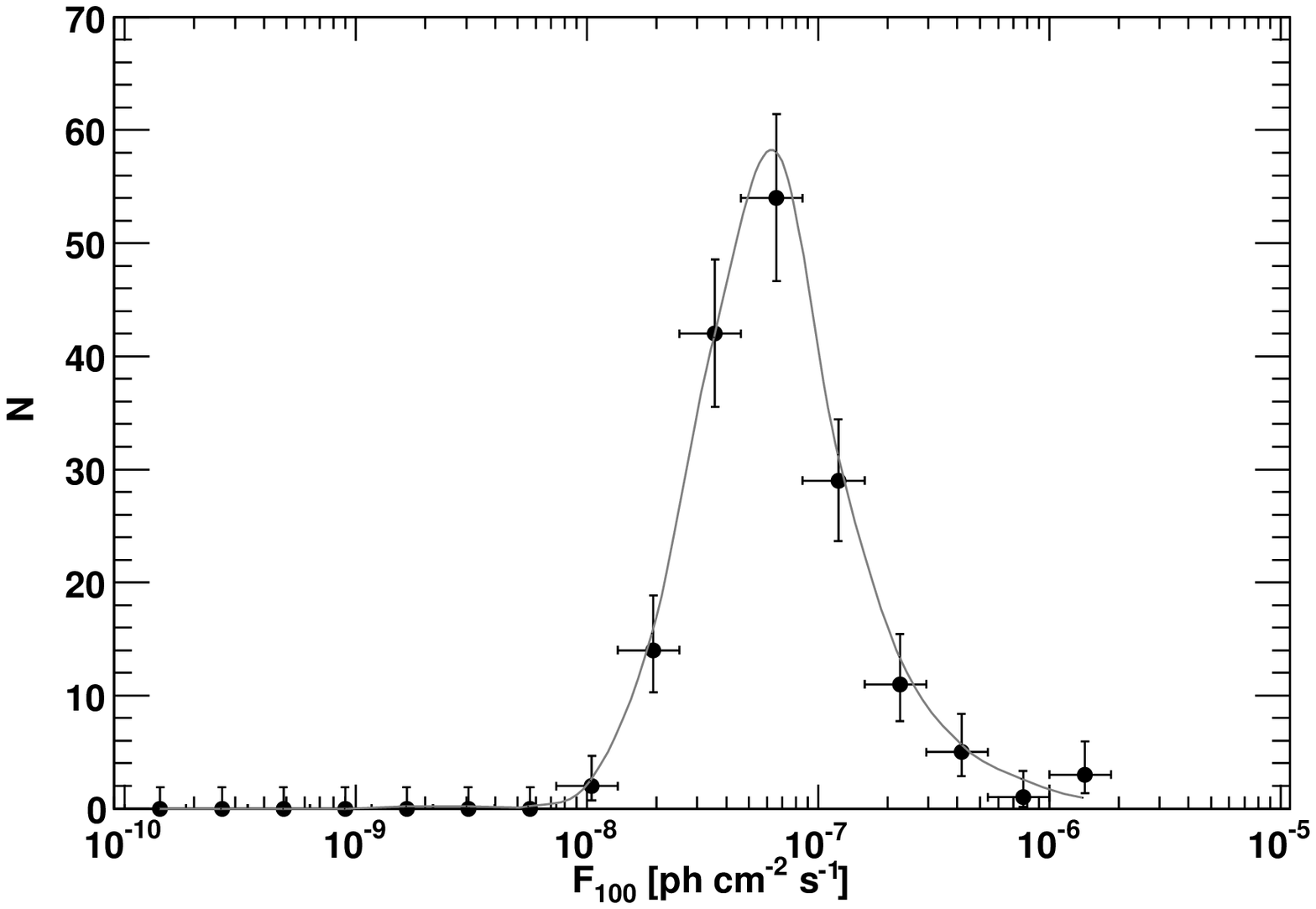}\\
\end{tabular}
  \end{center}
  \caption{Distribution of photon indices (left) and fluxes (right)
for the TS$\geq$50 and $|b|\geq$20$^{\circ}$ sources associated
with FSRQs. 
}
  \label{fig:logn_fsrq}
\end{figure*}

\begin{figure*}[ht!]
  \begin{center}
  \begin{tabular}{cc}
    \includegraphics[scale=0.43]{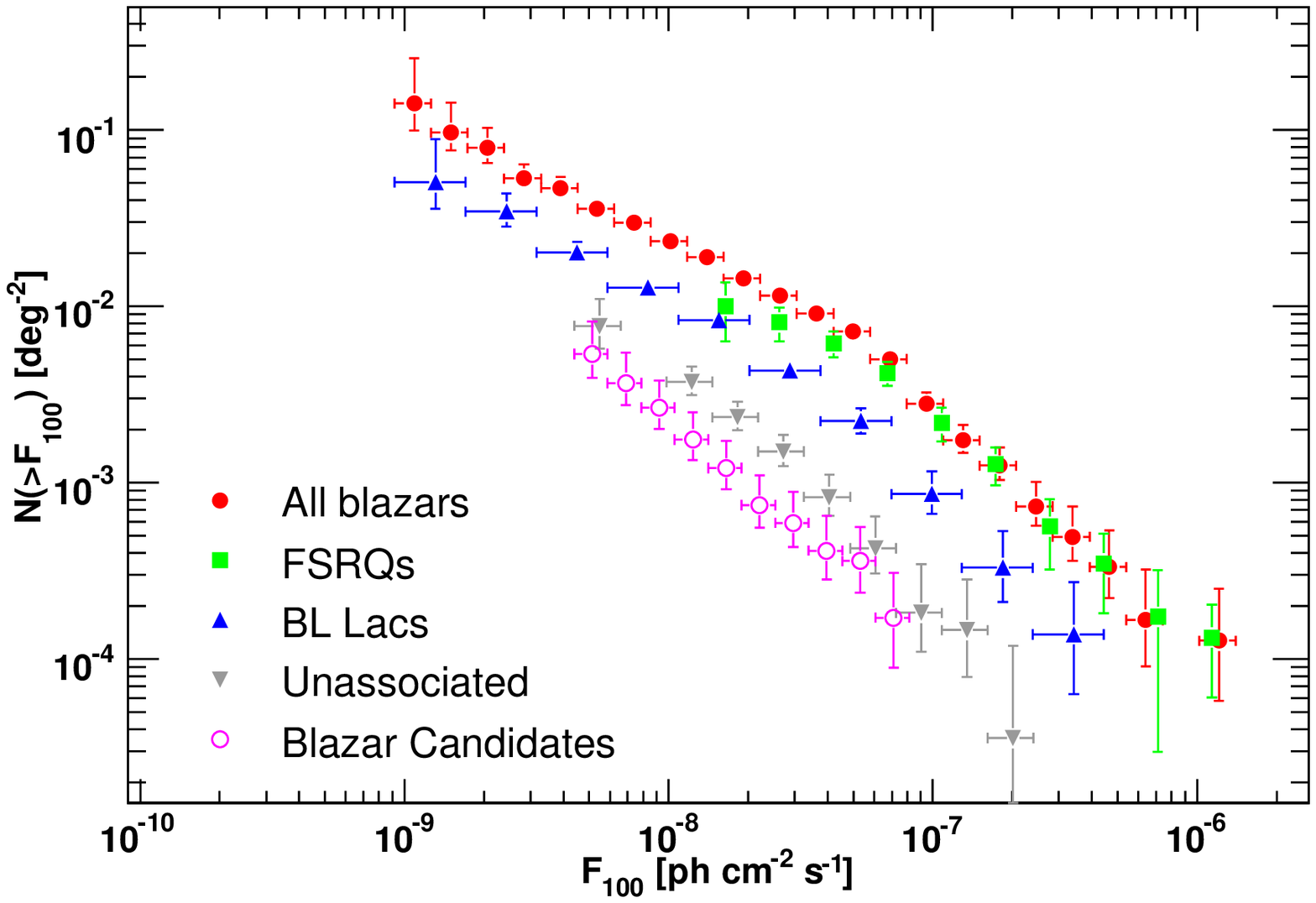} 
	 \includegraphics[scale=0.43]{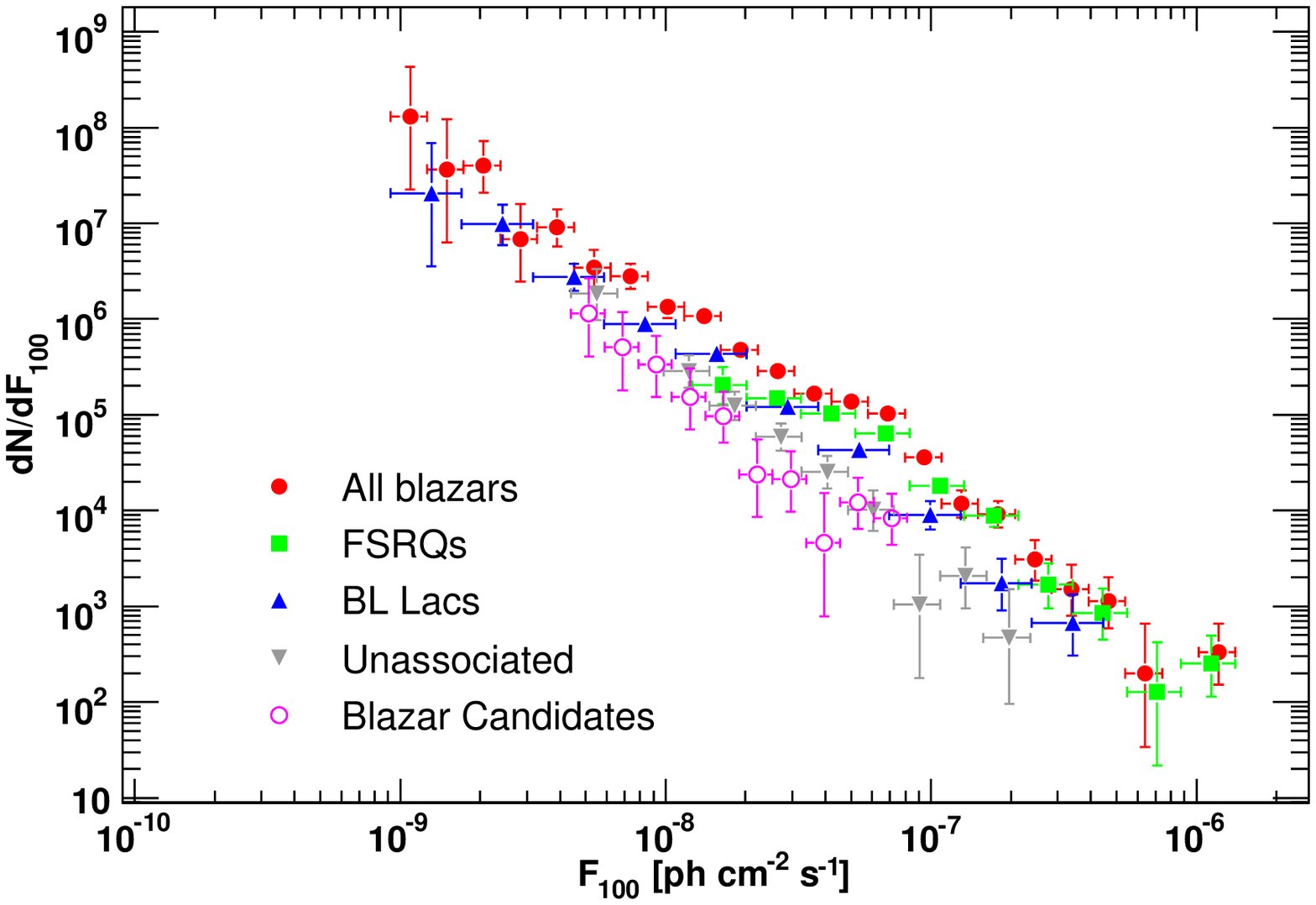}\\
\end{tabular}
  \end{center}
  \caption{
Cumulative (left) and differential (right) source count distribution
of {\it Fermi} blazars and the sub-samples reported in Tab.~\ref{tab:logn2D}.
Given the selection effect towards spectrally hard
sources, BL Lac objects are detected to fluxes fainter than FSRQs. The 
flattening at low fluxes of the FSRQs log$N$--log$S$ is probably due to 
incompleteness (see text for details). The ``All Blazars'' class also includes
all those sources which are classified as blazar candidates 
(see Tab.~\ref{tab:sample} for details).
}
  \label{fig:blazar_all}
\end{figure*}

The best fit to the source counts (reported in 
Tab.~\ref{tab:logn2D}) is a double power-law
model with a bright-end slope of 2.41$\pm0.16$ and faint-end slope
0.70$\pm0.30$. The log $N$--log $S$ relationship shows a break around 
F$_{100}=$6.12($\pm1.30)\times10^{-8}$\,ph cm$^{-2}$ s$^{-1}$.
The intrinsic distribution of the photon indices of FSRQs is found
to be compatible with a Gaussian distribution with mean and  dispersion
of 2.48$\pm0.02$ and 0.18$\pm0.01$ in agreement with what found
previously in Tab.~\ref{tab:index}. The faint-end slope is noticeably
flatter and this might be due to the fact that many of the unassociated
sources below the break might be FSRQs.
Fig.~\ref{fig:logn_fsrq} shows how the best-fit model reproduces 
the observed photon index and flux distributions.
The $\chi^2$ test indicates that the probability that the 
real distribution and the model line come from the same
parent population is $\geq0.99 $ for both
 the photon index and flux distributions respectively.
The left panel shows that the photon index distribution is not reproduced
perfectly. This might be due to incompleteness or by the
fact that the intrinsic distribution of photon indices is actually
not Gaussian. However, a Kolmogorov-Smirnov (KS) test between the
predicted and the observed distribution yields that  both
distributions have a probability of $\sim96$\,\% of being
drawn from the same parent population. Thus the current dataset
is compatible with the hypothesis that the intrinsic
index distribution is Gaussian.
The log $N$--log $S$ of FSRQs is shown in Fig.~\ref{fig:blazar_all}.


\subsection{BL Lacs}
\label{sec:bllac}

\begin{figure*}[ht!]
  \begin{center}
  \begin{tabular}{cc}
    \includegraphics[scale=0.4]{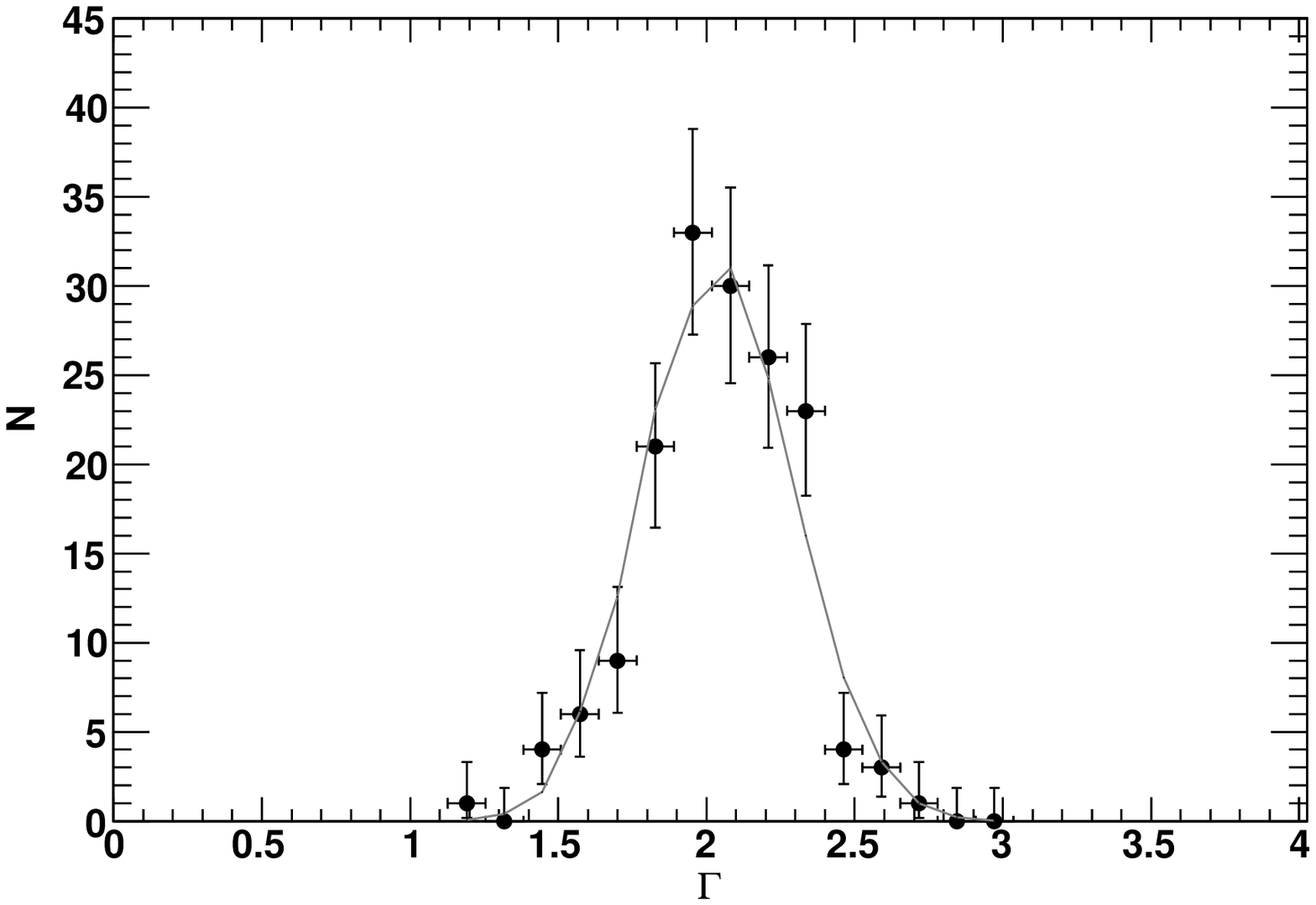} 
	 \includegraphics[scale=0.4]{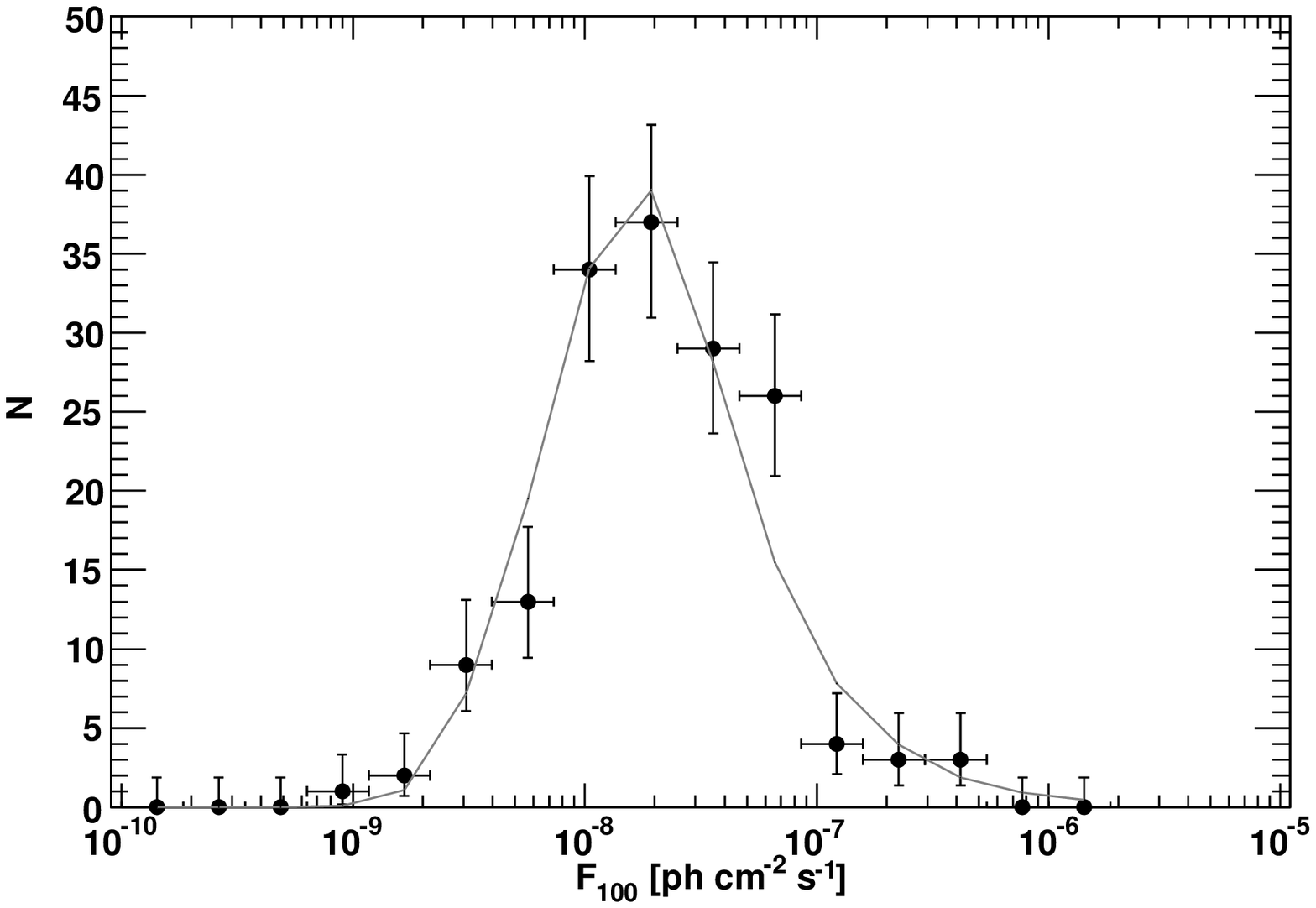}\\
\end{tabular}
  \end{center}
  \caption{Distribution of photon indices (left) and fluxes (right)
for the TS$\geq$50 and $|b|\geq$20$^{\circ}$ sources associated
with BL Lacs. The dashed line is the best fit $dN/dSd\Gamma$ model.
}
  \label{fig:logn_bllac}
\end{figure*}

The best-fit model of the source count distribution of the
161 BL Lac objects is again a broken power-law model.
The break is found to be at F$_{100}=$6.77$\pm1.30\times10^{-8}$\,ph cm$^{-2}$ s$^{-1}$ while  the  slopes below and above the break are  1.72$\pm0.14$ and
2.74$\pm0.30$ respectively.
The intrinsic photon index distribution is found
to be compatible with a Gaussian distribution with mean
and dispersion of 2.18$\pm0.02$ and  0.23$\pm0.01$ respectively.
These results are in good agreement with the one reported in 
Tab.~\ref{tab:index}. The best-fit parameters to the source
counts distribution are reported in Tab.~\ref{tab:logn2D}.
Fig.~\ref{fig:logn_bllac} shows how the best-fit model reproduces 
the observed photon index and flux distributions.
The $\chi^2$ test indicates that the probability that the 
real distribution and the model line come from the same
parent population is $\geq0.99$ for both
 the photon index and flux distributions respectively.
The log $N$--log $S$ of BL Lacs, compared to the one of FSRQs and blazars,
 is shown in Fig.~\ref{fig:blazar_all}.

\subsection{Unassociated Sources}
\label{sec:unids}
We also constructed the  log $N$--log $S$ of the 56 sources
which are unassociated and it is reported in Fig.~\ref{fig:blazar_all}.
Their source count distribution
displays a very steep bright-end slope ($\beta_1$=3.16$\pm0.50$),
a break around $\sim$4.5$\times10^{-8}$\,ph cm$^{-2}$ s$^{-1}$ and 
faint-end slope of  1.63$\pm0.24$. The intrinsic photon index
distribution is found to be compatible with a Gaussian distribution
with mean and dispersion of 2.29$\pm0.03$ and  0.20$\pm0.01$ respectively
(see Tab.~\ref{tab:logn2D} for details).
The extremely steep bright-end
slope is caused by the fact that most (but not all) of the
brightest sources have an association. Below the break the log $N$--log $S$
behaves like the one of blazars with the difference that the 
index distribution is suggesting that
probably most of the sources are BL Lac objects.
Indeed as  can be seen in Fig.~\ref{fig:blazar_all} all the
sources with F$_{100}\leq4\times10^{-8}$\,ph cm$^{-2}$ s$^{-1}$ are identified
as BL Lac objects in our sample.

\subsection{Unfolding Analysis}
\label{unf}
Finally we employ a different approach to evaluate the log $N$--log $S$ distribution based on a deconvolution (unfolding) technique. This method allows reconstructing the distribution of the number of sources from the data without assuming any model, also taking into account
the finite resolution (i.e. dispersion) of the sky coverage.

The purpose of the unfolding is to estimate the true distribution (cause) given the observed one (effect), and assuming some knowledge about the eventual migration effects (smearing matrix) as well as the efficiencies. The elements of the smearing  matrix represent the probabilities to observe a given effect that falls in an observed bin $Effect_j$ from a cause in a given true bin $Cause_i$. In our case the observed distribution represents the number of sources as function of the observed flux above 100\,MeV, while the true distribution represents the number of true sources as function of the true flux above 100\,MeV. The unfolding algorithm adopted here is based on the Bayes theorem \cite{dago}.

The smearing matrix is evaluated using the Monte Carlo simulation described in the $\S$~4. Its elements, $P(F100_{j,obs} | F100_{i,true})$, represent the probabilities that a source with a true flux above $100$\,MeV, $F100_{i,true}$, is reconstructed with an observed flux above $100$\,MeV, $F100_{j,obs}$. The data are assumed to be binned in histograms. The bin widths and the number of bins can be chosen independently for the distribution of the observed and reconstructed variables.

The log $N$--log $S$ reconstructed with this method is shown
in Fig.~\ref{fig:logn_all} and it is apparent that the source
counts distributions derived with the 3 different methods are 
all in good agreement with each other.

\subsection{Comparison with Previous Estimates}
\label{sec:comp}

\begin{figure}[h!]
\begin{centering}
	\includegraphics[scale=0.7]{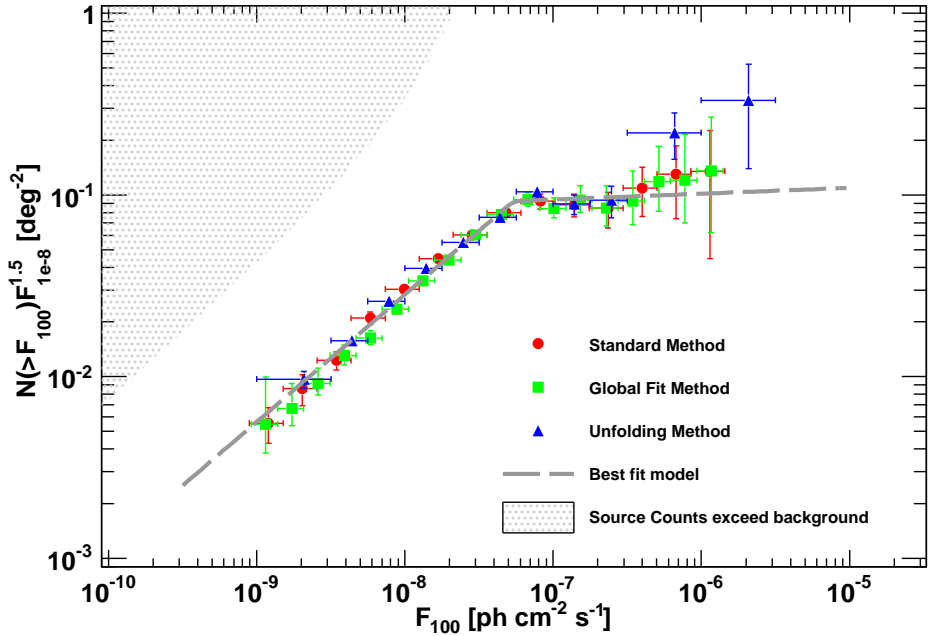} 
	\caption{Source count distribution of {\it Fermi} point-like sources derived
with three different methods. The distribution has been multiplied 
by (F$_{100}/10^{-8}$)$^{1.5}$. The dashed
line is the best fit model described in the text. The grey region
indicates the flux at which a power law connecting the log $N$--log $S$ break
(at $\sim6.6\times10^{-8}$\,ph cm$^{-2}$ s$^{-1}$) and that given flux
exceeds the EGB emission (see text for details).
}
	\label{fig:logn_all}
\end{centering}
\end{figure}

Fig.~\ref{fig:logn_all} shows that the  log $N$--log $S$ distributions
displays a strong break at fluxes F$_{100}\approx6\times10^{-8}$\, ph cm$^{-2}$
s$^{-1}$. This represents the first time that  such a flattening
is seen in the log $N$--log $S$ of $\gamma$-ray sources, blazar
in particular. This is due to the fact
that {\it Fermi} couples a good sensitivity to the all-sky coverage thus
allowing to determine the source counts distribution over more than
3 decades in flux.

Above fluxes of F$_{100}=10^{-9}$\,ph cm$^{-2}$ s$^{-1}$, the
surface density of sources is 0.12$^{+0.03}_{-0.02}$\, deg$^{-2}$.
At these faint fluxes our comparison can only be done with 
predictions from different models.
\cite{dermer07} and \cite{inoue09} predict a blazar surface density
of respectively 0.030\,deg$^{-2}$ 
and  0.033\,deg$^{-2}$. Both these predictions are a factor
$\sim4$ below the LAT measurement. However, it should be stressed that
these models are based on the EGRET blazar sample which, because of strong
selection effects against high-energy photons, counted 
a very limited number of BL Lac objects.

At brighter fluxes (F$_{100}\geq5\times10^{-8}$\,ph cm$^{-2}$ s$^{-1}$)
\cite{dermer07} predicts a density of FSRQs and BL Lacs of 
4.1$\times10^{-3}$\,deg$^{-2}$ and 1.1$\times10^{-3}$\,deg$^{-2}$ respectively.
At the same flux, \cite{muecke00} predict a density of 
1.21$\times10^{-3}$\,deg$^{-2}$ and 3.04$\times10^{-4}$\,deg$^{-2}$ respectively
for FSRQs and BL Lac objects.
The densities measured by {\it Fermi} are significantly larger, 
being  6.0$(\pm0.6)\times10^{-3}$\,deg$^{-2}$ for FSRQs
and 2.0$(\pm 0.3)\times10^{-3}$\,deg$^{-2}$ for BL Lacs.

\section{Analysis in the Different Energy Bands}
\label{sec:bands}
The aim of the following analysis  is to determine the contribution
of point sources to the EGB in different contiguous energy bands.
This is done by creating a log$N$--log$S$ distribution in 3 different
energy bands: 
0.1--1.0\,GeV, 1.0--10.0\,GeV and 10.0--100\,GeV bands.
This will allow us to study 
the spectrum of the unresolved emission from point sources
and at the same time explore the properties
of the source population in different bands. With this approach,
the systematic uncertainty related to the flux estimate,
given by the complex spectra of blazars (see $\$$~\ref{sec:pow}), 
will be removed.
In addition, use of these bands should allow
us to extend the survey region to $|b|\geq10^{\circ}$ 
(see $\S$~\ref{sec:mlfit}).

The analysis follows the method outlined in $\S$~\ref{sec:sim} with
the difference that the final ML fit is restricted to the band
under investigation. In the spectral fit, all parameters (including
the photon index) are left free and are optimized by maximizing the likelihood
function.
Only sources that a given band have TS$\geq$25 are considered
detected in that band. Formally each band and related sample is treated
as independent here and no prior knowledge of the source spectral behaviour
is assumed. In the three bands, the samples comprise respectively
362, 597 and 200 sources detected for $|b|\geq$10$^{\circ}$ and TS$\geq25$.

\begin{figure*}[ht!]
  \begin{center}
  \begin{tabular}{ccc}
  \includegraphics[scale=0.3]{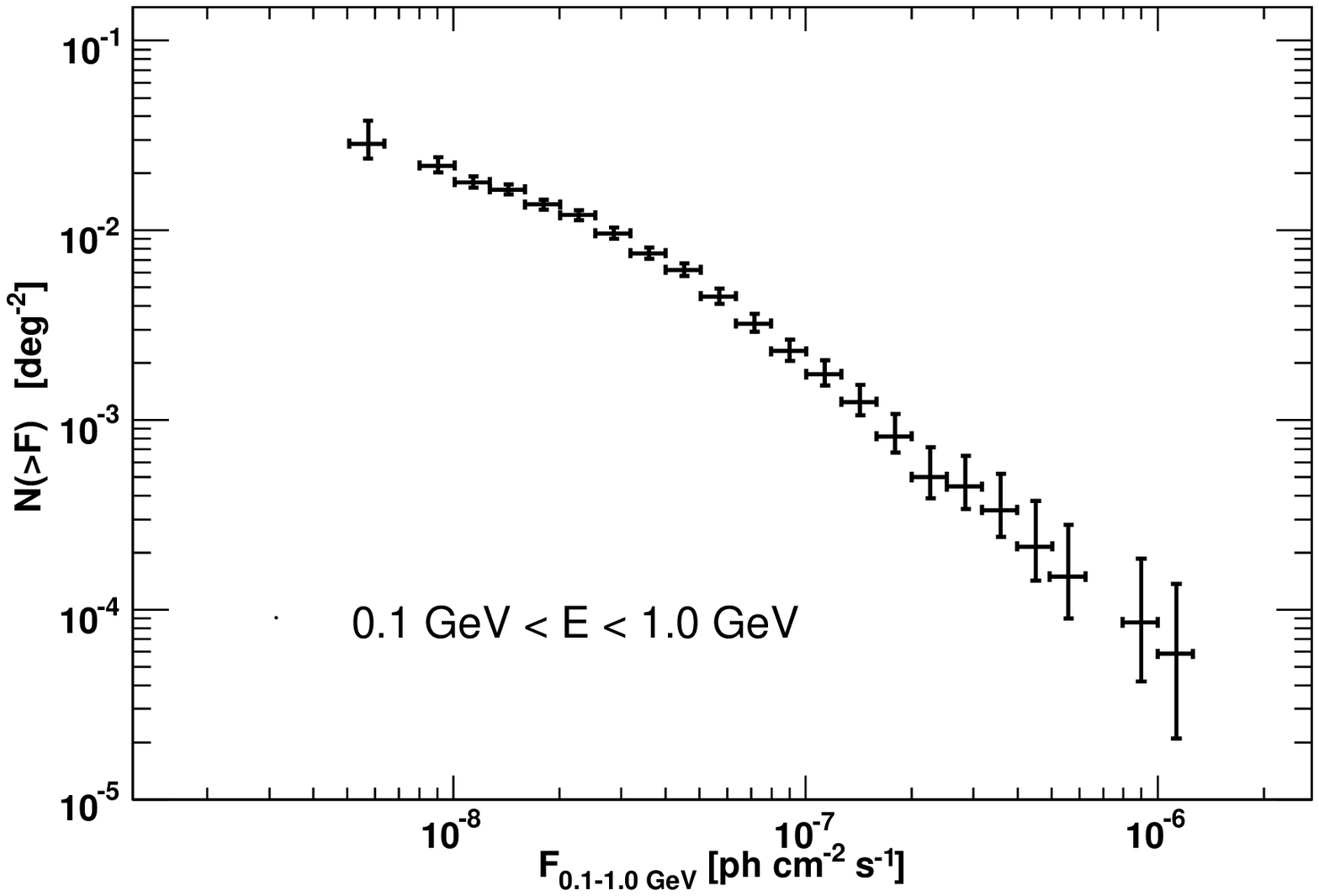} 
	 \includegraphics[scale=0.3]{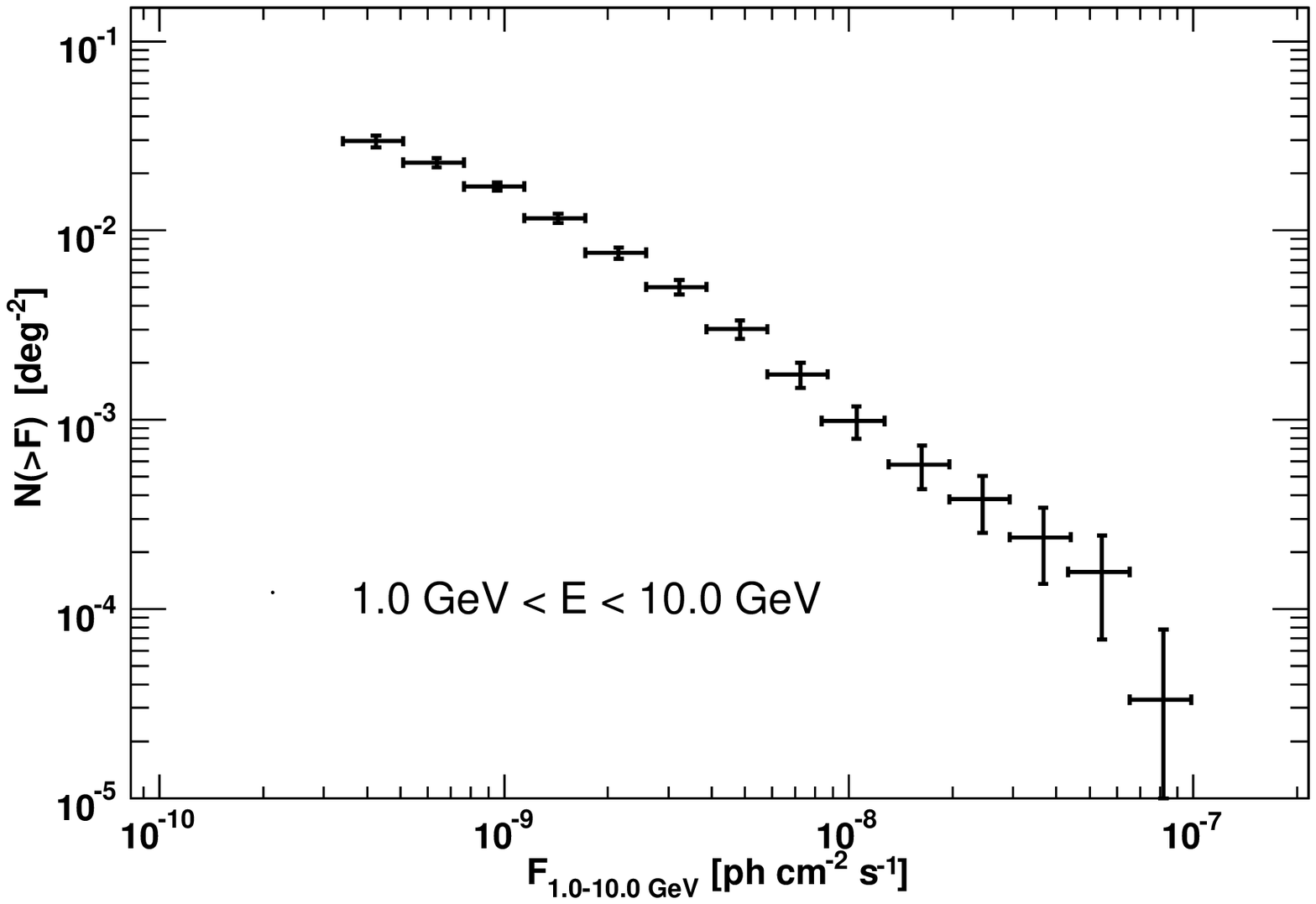}
  \includegraphics[scale=0.3]{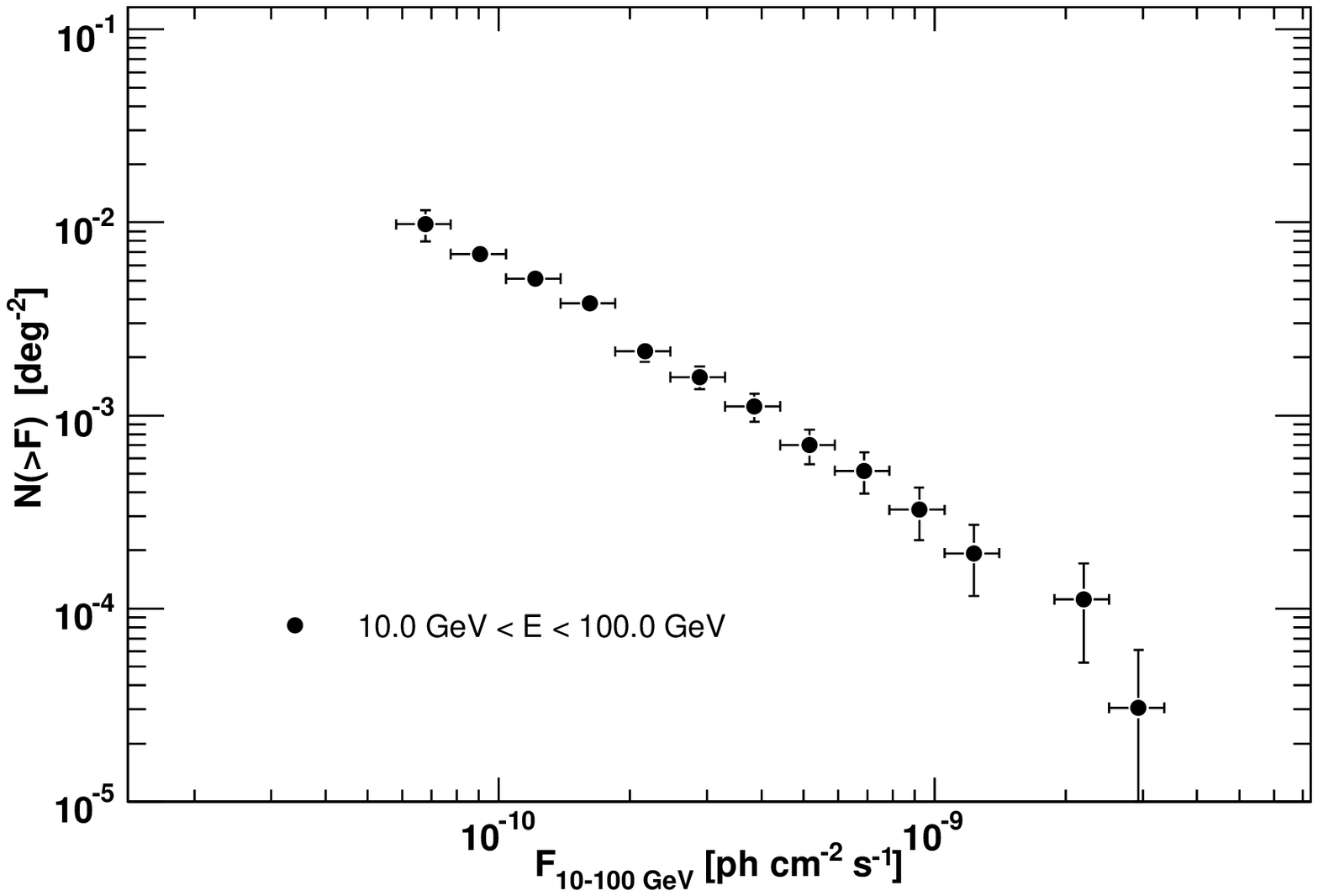} 
\end{tabular}
  \end{center}
  \caption{Source count distributions for the soft (0.1-1.0\,GeV, left),
medium (1.0--10.0\,GeV, center) and high energy (10.0--100.0\,GeV, right)
band reconstructed with the method reported in $\S$~\ref{sec:logn_2d}.
}
  \label{fig:logn_bands}
\end{figure*}

In both the soft and the medium band (i.e. 0.1--1.0\,GeV and 1.0--10.0\,GeV), 
the log$N$--log$S$ is well described by a double power-law model, while
in the hardest band (10--100\,GeV) the  log$N$--log$S$
is compatible with a single power-law model with a differential slope
of 2.36$\pm0.07$. The results of the best-fit models are reported in
Tab.~\ref{tab:logn_bands} and are shown in Fig.~\ref{fig:logn_bands}.
The {\it spectral bias} (see $\S$~\ref{sec:term})
is the strongest in the soft band while it is absent in the high-energy band,
being already almost negligible above 1\,GeV.

From the log$N$--log$S$ in the whole band we would expect 
(assuming a power law with a photon index of 2.4 and that
the blazar population is not changing dramatically with energy) to 
find breaks at: 6.7$\times10^{-8}$,  2.6$\times10^{-9}$, and
1$\times10^{-10}$ ph cm$^{-2}$ s$^{-1}$ for the soft, medium, and hard bands
respectively. Indeed these expectations are confirmed by the ML fits
in the individual bands (e.g. see Tab.~\ref{tab:logn_bands}).
The hard band constitutes the only exception where
the flux distribution barely extends below the flux at which the break
might be observed.

The average spectral properties of the sample change with energy.
We find that the {\it intrinsic} index distribution is compatible
with a Gaussian distribution with means of 2.25$\pm0.02$, 2.43$\pm0.03$,
and 2.17$\pm0.05$. In the three bands the fraction of BL Lac-to-FSRQ is:
0.61, 1.14, and 3.53 respectively with identification
incompletenesses of 0.18, 0.25, and 0.25 respectively.
It is apparent that the hardest band is the best one for studying
BL Lac objects since the contamination due to FSRQs is rather small.

%
%

\begin{deluxetable}{lccccccccccc}
\tablewidth{0pt}
\rotate
\tabletypesize{\scriptsize}

%
%
\tablecaption{
Results of the best fits to the source counts
distributions in different Energy bands. Parameters without an error
estimate were kept  fixed during the fitting stage.
\label{tab:logn_bands}}
\tablehead{
\colhead{} & \colhead{} & \colhead{} &
 \multicolumn{2}{c}{Sample Limits} & \colhead{} &
\multicolumn{6}{c}{Best-fit Parameters}\\
\cline{4-5} \cline{7-12} \\ 
\colhead{BAND} & \colhead{\# Objects } & \colhead{Incompl.} & 
\colhead{TS$\geq$}     & \colhead{$|$b$|\geq$} & \colhead{} &
\colhead{A\tablenotemark{a}} &  \colhead{$\beta_1$} &
\colhead{S$_b$\tablenotemark{b}}  & \colhead{$\beta_2$} &
\colhead{$\mu$}   &  \colhead{$\sigma$}
}

\startdata
0.1--1.0\,GeV & 362 & 0 & 25 & 10$^{\circ}$ & &4.00$\pm0.21$& 2.55$^{+0.17}_{-0.22}$ & 5.75$^{+0.44}_{-2.22}$ & 1.38$^{+0.13}_{-0.46}$ & 2.25$^{+0.02}_{-0.02}$ & 0.32$^{+0.01}_{-0.01}$ \\
1.0--10.0\,GeV & 597 &  0 & 25 & 10$^{\circ}$ & & 1.097$\pm0.05$ & 2.38$^{+0.15}_{-0.14}$ & 0.23$\pm0.06$ & 1.52$^{+0.8}_{-1.1}$ & 2.43 & 0.40$^{+0.02}_{-0.02}$\\
10.0--100.0\,GeV & 200 & 0 & 25 & 10$^{\circ}$ & &  8.3($\pm0.6$)$\times10^{-3}$ & 2.364$^{+0.07}_{-0.07}$ & \nodata &\nodata & 2.17$\pm0.05$ & 0.82$^{+0.05}_{-0.05}$\\
0.3--100.0\,GeV & 759 & 0 & 25 & 10$^{\circ}$ & &  5.33$\pm0.19$ & 2.44$^{+0.15}_{-0.11}$ & 1.69$^{+0.33}_{-0.33}$ & 1.70$^{+0.06}_{-0.07}$ & 2.35$^{+0.02}_{-0.02}$ & 0.30$^{+0.01}_{-0.01}$

\enddata
\tablenotetext{a}{In units of 10$^{-14}$\,cm$^{2}$ s deg$^{-2}$.}
\tablenotetext{b}{In units of 10$^{-8}$\,ph cm$^{-2}$ s$^{-1}$.}

\end{deluxetable}

\section{Contribution of Sources to the Diffuse Background}
\label{sec:edb}
The source count distribution can be used to estimate the contribution
of point-like sources to the EGB emission. This allows us to
determine the fraction of the GeV diffuse emission that 
arises from point-like source populations measured by {\it Fermi}.
As specified in $\S$~\ref{sec:term},
this estimate does not include the contribution of sources which
have been directly detected by {\it Fermi} since these are not considered
in the measurement of the diffuse background. This estimate includes all
those sources which, because the detection efficiency changes with flux,
photon index and position in the sky, have not been detected.

The diffuse emission arising from a class of sources can be determined as:
\begin{equation}
F_{\rm diffuse} = \int^{S_{\rm max}}_{S_{\rm min}}dS \int^{\Gamma_{\rm max}}_{\Gamma_{\rm min}} d\Gamma \frac{dN}{dSd\Gamma} 
S \left ( 1-\frac{\Omega(\Gamma,S)} {\Omega_{\rm max}} \right )
\label{eq:diff}
\end{equation}

where $\Omega_{\rm max}$ is the geometrical sky area and the 
$(1-\Omega(\Gamma,S)/\Omega_{\rm max})$ term takes into account that
the threshold at which LAT detects sources depends on both the
photon index and the source flux. We note that neglecting
the dependence of $\Omega$ on the photon index (i.e. using the 
mono-dimensional sky coverage reported in Fig.~\ref{fig:skycov})
would result in an underestimate of the diffuse flux resolved by {\it Fermi}
into point-sources. The limits of integration of Eq.~\ref{eq:diff}
are $\Gamma_{\rm min}=1.0$, $\Gamma_{\rm max}=3.5$, and $S_{\rm max}=10^{-3}$\,ph cm$^{-2}$ s$^{-1}$. We also note that  the integrand of 
Eq.~\ref{eq:diff} goes to zero for bright fluxes or for photon indices
which are either very small or very large; thus the integration
is almost independent of  the parameters reported above.
The integration is not independent of the value of $S_{\rm min}$ which
is set to the flux of the faintest  source detected in the sample.
For the analysis of the whole band $S_{\rm min}$=9.36$\times10^{-10}$\,ph cm$^{-2}$ s$^{-1}$ while for the low, medium and hard band S$_{\rm min}$ is set
to: 
5.17$\times10^{-9}$\,ph cm$^{-2}$ s$^{-1}$, 
3.58$\times10^{-10}$\,ph cm$^{-2}$ s$^{-1}$, and
6.11$\times10^{-11}$\,ph cm$^{-2}$ s$^{-1}$ respectively.

Since in the measurement of \cite{lat_edb} the sources which are subtracted
are those detected in 9\,months of operation, the coverage used
in Eq.~\ref{eq:diff} is  the one corresponding to the 9\,months
survey. The uncertainties on the diffuse flux have been computed by
performing a bootstrap analysis.
Integrating Eq.~\ref{eq:diff} 
we find that the point source contribution is
1.63$(\pm0.18)\times10^{-6}$\,ph cm$^{-2}$ s$^{-1}$ sr$^{-1}$ 
where  the systematic uncertainty is 0.6$\times10^{-6}$\,ph cm$^{-2}$ s$^{-1}$ sr$^{-1}$. 
This corresponds to 16$(\pm1.8)$\,\% ($\pm$7\,\% systematic uncertainty)
of the Isotropic diffuse emission
measured by LAT \citep{lat_edb} above 100\,MeV. This small fraction is a natural
consequence of the break of the source counts distribution.
However, it is also possible to show that the parameter space for
the faint-end slope $\beta_2$ is rather limited and that a break
must exist in the range of fluxes spanned by this analysis.
Indeed, for a given $\beta_2$ (and all the other parameters
of the log $N$--log $S$ fixed at their best-fit values)  
one can solve Eq.~\ref{eq:diff}
to determine the flux at which the integrated emission
of point sources exceeds the one of the EGB. Repeating
this exercise for many different values of the $\beta_2$ parameter yields
an exclusion region which constrains the behavior of the log $N$--log $S$
at low fluxes. The results of this exercise are shown in Fig.~\ref{fig:logn_all}.
From this Figure it is apparent that the log $N$--log $S$ {\it must}  break
between F$_{100}\approx2\times10^{-9}$\,ph cm$^{2}$ s$^{-1}$ and
F$_{100}\approx6.6\times10^{-8}$\,ph cm$^{2}$ s$^{-1}$.
For a small break (e.g. $\beta_1-\beta_2\approx 0.2-0.3$ and then 
$\beta_2\approx$2.2--2.3), the integrated emission of point sources
would still match the intensity of the diffuse background at
F$_{100}\approx 10^{-9}$\,ph cm$^{2}$ s$^{-1}$ which are sampled
by {\it Fermi}. Thus not only the break has to exist, but this
simple analysis shows that it has to be strong (see also $\S$~\ref{sec:siml}),
not to exceed the intensity of the diffuse emission.

\begin{figure*}[ht!]
  \begin{center}
  \begin{tabular}{c}
  \includegraphics[scale=0.7]{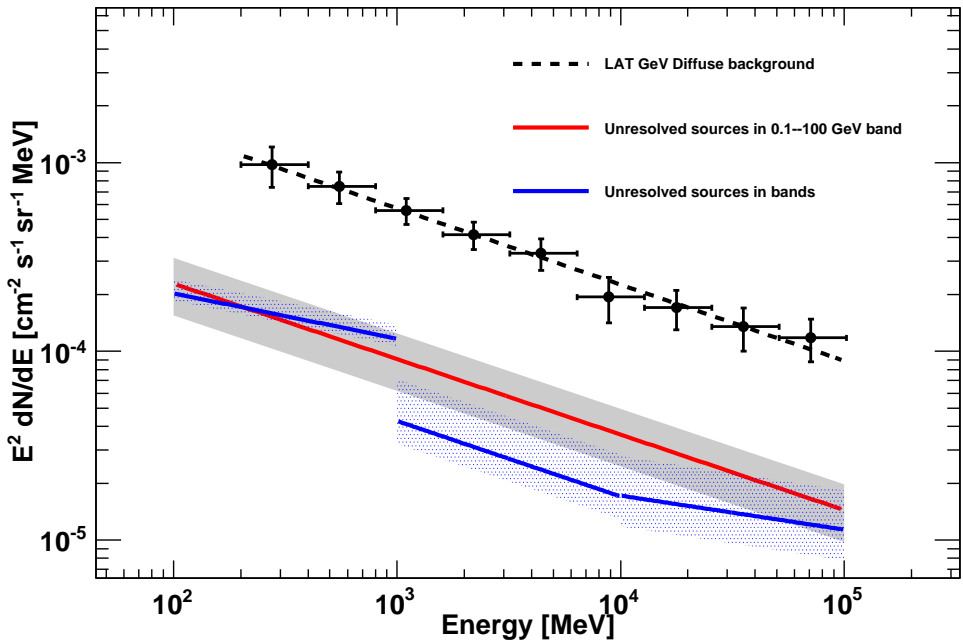} 
\end{tabular}
  \end{center}
  \caption{Contribution of point-sources to the diffuse GeV background.
The red solid line was derived from the study of the log$N$--log$S$ in the
whole band while the blue solid lines come from the study of individual
energy bands (see $\S$~\ref{sec:bands}). The bands (grey solid and hatched
blue) show the total (statistical plus systematic) uncertainty.
}
  \label{fig:edb}
\end{figure*}

The log$N$--log$S$ in the whole band goes deeper than the source count
distributions derived in the smaller bands. This is clearly shown
in Fig.~\ref{fig:edb}. Given the fact that most of the source
flux is emitted below 1\,GeV (for reasonable photon indices),
the source count distribution in the soft band (0.1--1.0\,GeV)
is the one which gets closer to the log$N$--log$S$ in the whole band
in terms of resolved diffuse flux.

\begin{figure*}[ht!]
  \begin{center}
  \begin{tabular}{c}
  \includegraphics[scale=0.7]{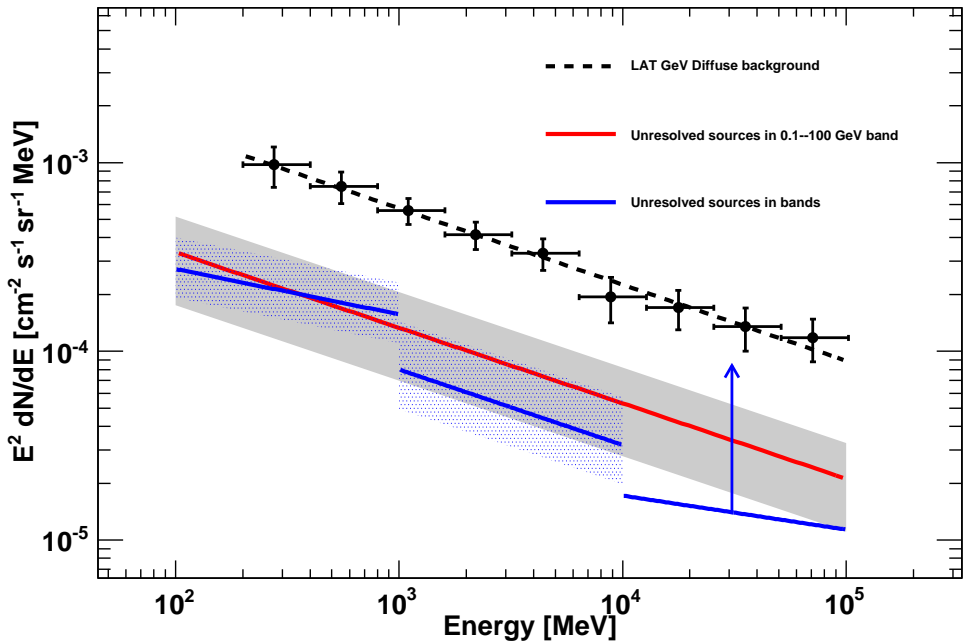} 
\end{tabular}
  \end{center}
  \caption{Contribution of point-sources to the diffuse GeV background
obtained by extrapolating and integrating the log $N$--log $S$
to zero flux.
The red solid line was derived from the study of the log$N$--log$S$ in the
whole band while the blue solid lines come from the study of individual
energy bands (see $\S$~\ref{sec:bands}). The bands (grey solid and hatched
blue) show the total (statistical plus systematic) uncertainty.
The arrow indicates the lower limit on the integration of Eq.~\ref{eq:diff}
for the 10--100\,GeV band.
}
  \label{fig:edb2}
\end{figure*}

The log $N$--log $S$ in the whole bands 
shows a strong break with a faint-end slope (e.g. $\beta_2$) robustly
constrained to be $<$2. In this case the integral reported
in Eq.~\ref{eq:diff} converges for small fluxes and it can be evaluated
at zero flux to assess the
maximum contribution of {\it Fermi}-like sources to the diffuse background.
This turns out to be 
2.39$(\pm0.48)\times10^{-6}$\,ph cm$^{-2}$ s$^{-1}$ sr$^{-1}$ 
(1.26$\times10^{-6}$\,ph cm$^{-2}$ s$^{-1}$ sr$^{-1}$ systematic uncertainty)
which represents 23$(\pm5)$\,\% (12\,\% systematic uncertainty) 
of the {\it Fermi} diffuse background \citep{lat_edb}. 
This is a correct result as long as the   log $N$--log $S$
of point-sources (i.e. blazars) does not become steeper at fluxes below
the ones currently sampled by {\it Fermi}. A given source population normally exhibits
a source count distribution with a single downwards break  \citep[e.g. see
the case of radio-quiet AGN in][]{cappelluti07}. This break is 
of cosmological origin since it coincides with the change of sign of the
evolution of that population.
As can be clearly seen in the redshift distribution 
in \cite{agn_cat}  the epoch of maximum
growth of blazars corresponds to redshift 1.5--2.0 which
coincides well with the peak of the star formation in the Universe
\citep[e.g.][]{hopkins06}. Since {\it Fermi} is already sampling this population
it is reasonable to expect  no other breaks  in the source count distribution
of blazars. Under this assumption, the result of the integration of 
Eq.~\ref{eq:diff}  are correct. The results of this exercise are 
shown in Fig.~\ref{fig:edb2} and summarized in Tab.~\ref{tab:diffuse}.
Since the 10--100\,GeV source counts distribution 
does not show a break, its integral diverges for small fluxes.
Thus, in both Fig.~\ref{fig:edb2} and Tab.~\ref{tab:diffuse}
we decided to adopt, as a lower limit to the contribution of sources
to the diffuse emission in this band, the value of the integral
evaluated at the flux of the faintest detected source.

\begin{deluxetable}{lcccc}
\tablecolumns{5} 
\tablewidth{0pc}
\tabletypesize{\footnotesize}
\tablecaption{Diffuse emission arising from point sources. The lower part
of the table shows the values of the integrated emission when
the source counts distributions are extrapolated to zero flux.
Errors are statiscal only (see $\S$~\ref{sec:syst} for a discussion
about systematic uncertainties).
\label{tab:diffuse}}
\tablehead{
\colhead{Band} &   \colhead{EGB Intensity\tablenotemark{a}}   & 
\colhead{Point Source Diffuse Emission}      & 
\colhead{Fraction of EGB Intensity}     &
\colhead{S$_{min}$\tablenotemark{b}}    \\                      

\colhead{\scriptsize (GeV)} & \colhead{\scriptsize(ph cm$^{-2}$ s$^{-1}$ sr$^{-1}$)} &
\colhead{\scriptsize(ph cm$^{-2}$ s$^{-1}$ sr$^{-1}$)} &
\colhead{\scriptsize(\%)} &
\colhead{\scriptsize (ph cm$^{-2}$ s$^{-1}$)} 
}
\startdata
0.1--100 & 1.03$\times10^{-5}$ & 1.63$(\pm0.18)\times10^{-6}$ &
15.8($\pm1.8$) & 9.36\\
0.1--1.0 & 9.89$\times10^{-6}$ & 1.54$^{+0.29}_{-0.13}\times10^{-6}$& 
15.5$^{+2.9}_{-1.3}$ & 51.1\\
1.0--10 & 3.85$\times10^{-7}$  & 2.93$^{+1.95}_{-0.71}\times10^{-8}$& 
7.6$^{+5.0}_{-1.8}$ & 3.58 \\
10--100 & 1.50$\times10^{-8}$ & 1.36$^{+0.84}_{-0.43}\times 10^{-9}$ &
9.0$^{+5.6}_{-2.8}$ & 0.61\\
\hline
0.1--100 & 1.03$\times10^{-5}$ & 2.39$(\pm0.48)\times10^{-6}$ &
22.5($\pm1.8$) & 0\\
0.1--1.0 & 9.89$\times10^{-6}$ & 2.07$^{+0.98}_{-0.61}\times10^{-6}$ &
20.9$^{+10.0}_{-6.1}$ & 0\\
1.0--10 & 3.85$\times10^{-7}$ & 5.49$^{+4.36}_{-2.10}\times10^{-8}$ &
14.2$^{+11.2}_{-5.4}$ & 0\\
10--100 & 1.50$\times10^{-8}$ & $>1.36\times10^{-6}$& $>9.0$  &
0\tablenotemark{c}\\
\enddata
\tablenotetext{a}{The intentisities of the EGB emission are
derived from \cite{lat_edb}.} 
\tablenotetext{b}{\ Lower flux of integration of the source counts
distributions (see Eq.~\ref{eq:diff}) in units of $10^{-10}$\,ph cm$^{-2}$
s$^{-1}$.}
\tablenotetext{c}{The source counts distribution in the 10--100\,GeV
does not show a break and thus, its integral to zero flux diverges.
As a lower limit on the diffuse emission, we
 adopted the value computed at the faintest detected source.  }

\end{deluxetable}

\begin{figure*}[ht!]
  \begin{center}
  \begin{tabular}{c}
  \includegraphics[scale=0.7]{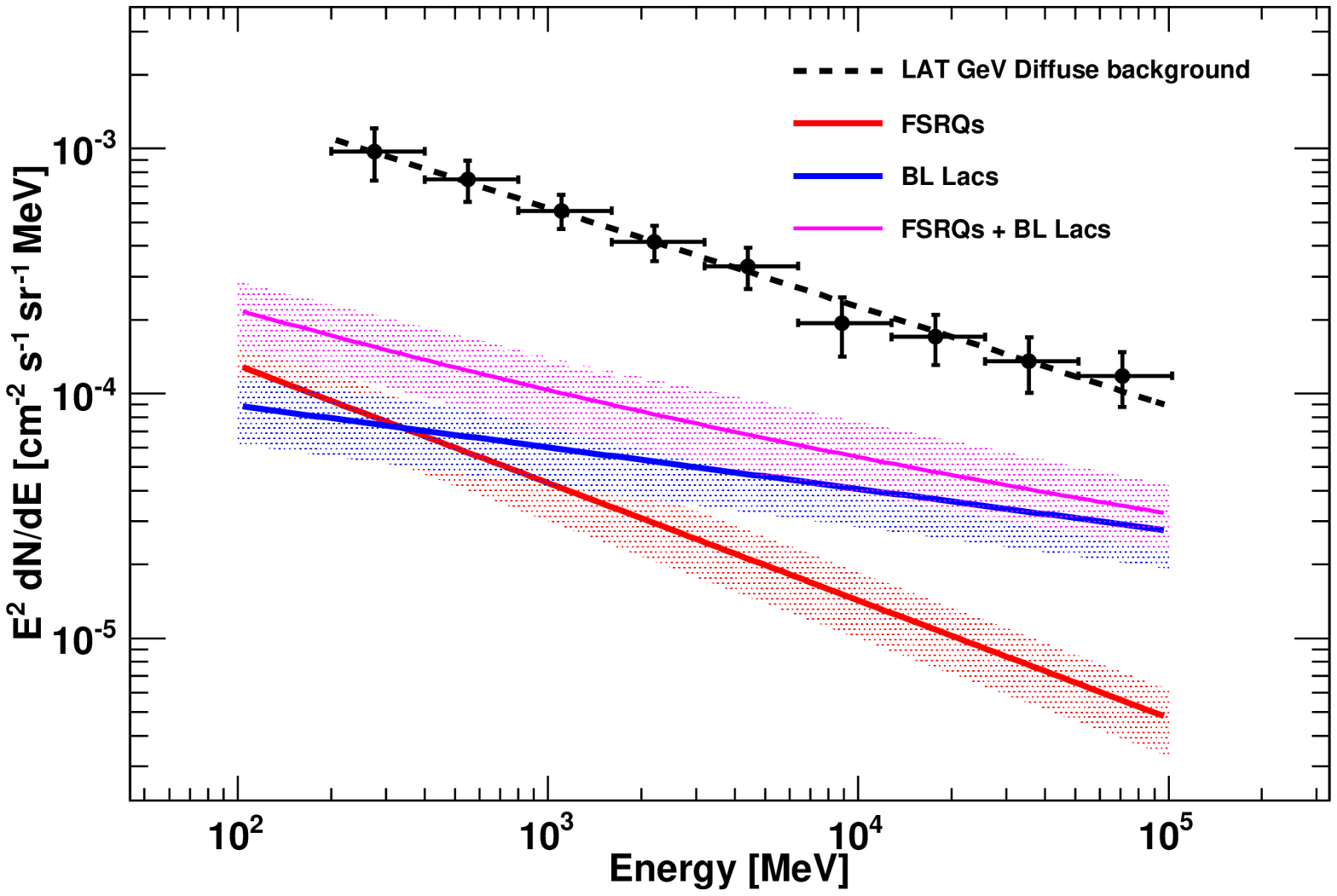} 
\end{tabular}
  \end{center}
  \caption{Contributions of different classes of blazars to the 
diffuse GeV background obtained by integrating the log $N$--log $S$.
The red and the blues solid lines show the contribution of FSRQs and 
BL Lacs respectively, while the pink solid line shows the sum of the two.
The bands around each line  
show the total (statistical plus systematic) uncertainty.
}
  \label{fig:edb3}
\end{figure*}

The different levels of contribution to the diffuse background as a function
of energy band might be the effect of the mixing of the two blazar populations.
In other words, as shown in $\S$~\ref{sec:bands}, FSRQs are the dominant
population below 1\,GeV while BL Lacs are the dominant one above 10\,GeV.
Given also that FSRQs are softer than BL Lacs 
(see also $\S$~\ref{sec:spec}), it is 
naturally to expect  a modulation in the blazar
contribution to the diffuse emission as a function of energy.
This can clearly be seen in Fig.~\ref{fig:edb3} which shows
the contribution of FSRQs and BL Lacs to the diffuse emission.
This has been computed integrating the source count distribution
of Tab.~\ref{tab:logn2D} to the minimum detected source flux
which is 9.36$\times10^{-10}$\,ph cm$^{-2}$ s$^{-1}$ and
and 1.11$\times10^{-8}$\,ph cm$^{-2}$ s$^{-1}$ for BL Lacs and FSRQs
respectively. It is clear that FSRQs are contributing most of the 
blazar diffuse emission below 1\,GeV while BL Lacs, given their hard
spectra, dominate above a few GeVs. The spectrum of the diffuse emission
arising from the blazar class is thus curved, being soft at low energy (e.g.
below 1\,GeV) and hard at high energy (above 10\,GeV), in agreement
with the results of the analysis of the source count distributions
in different bands.

\subsection{Additional Tests}

\subsection{Source Count Distribution above 300\,MeV}
\begin{figure*}[ht!]
  \begin{center}
  \begin{tabular}{c}
  \includegraphics[scale=0.7]{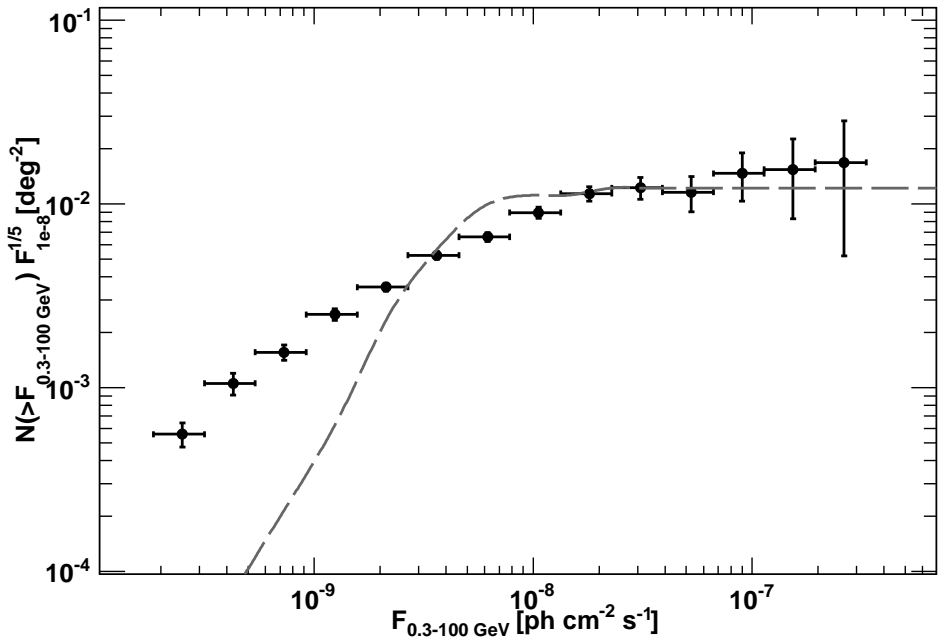} 
\end{tabular}
  \end{center}
  \caption{Source count distribution of all (TS$\geq25$, $|b|\geq10^{\circ}$)
sources in the 300\,MeV--100\,GeV band. The distribution has been multiplied 
by (F$_{100}/10^{-8}$)$^{1.5}$. The dashed line shows the sky coverage 
(scaled by an arbitrary factor) used to derive the source counts.
Note that the break of the log $N$ -- log $S$ and that one of the sky coverage
are at different fluxes.
}
  \label{fig:logn300mev}
\end{figure*}

The effective area of the LAT decreases quickly below 300\,MeV
while at the same time both the PSF size and the intensity of the 
diffuse background increase \citep[e.g. see ][]{atwood09}. 
In particular at the lowest energies,
systematic uncertainties in the instrument response might compromise
the result of the maximum likelihood fit to a given source (or
set of sources). In order to overcome this limitation we constructed,
with the method outlined in $\S$~\ref{sec:bands},
the log $N$--log $S$ of point sources in the 300\,MeV--100\,GeV band.
Considering that  in the E$>100$\,MeV band the log $N$--log $S$
shows a break around 6-7$\times10^{-8}$\,ph cm$^{-2}$ s$^{-1}$ and
assuming a power law with a photon index of 2.4, we would
expect to detect a break in the (E$\geq$300\,MeV) log $N$--log $S$ around 
$\sim$1.5$\times10^{-8}$\,ph cm$^{-2}$ s$^{-1}$. Indeed,
as shown in Fig.~\ref{fig:logn300mev}, the break is detected at
1.68$(\pm0.33)\times10^{-8}$\,ph cm$^{-2}$ s$^{-1}$.
Moreover, as Fig.~\ref{fig:logn300mev} shows, the break of the log $N$--log $S$
and the one of the sky coverage are at  different fluxes.
More precisely, the source counts start to bend down before the sky coverage
does it. This is an additional confirmation, along with the results of 
$\S$~\ref{sec:bands}, that the break of the log $N$--log $S$ is not
caused by the sky coverage. The parameters of this additional
source count distribution are reported for reference in 
Tab.~\ref{tab:logn_bands}.

\subsection{Simulating a log $N$--log $S$ without a break}
\label{sec:siml}
\begin{figure}[h!]
\begin{centering}
	\includegraphics[scale=0.6]{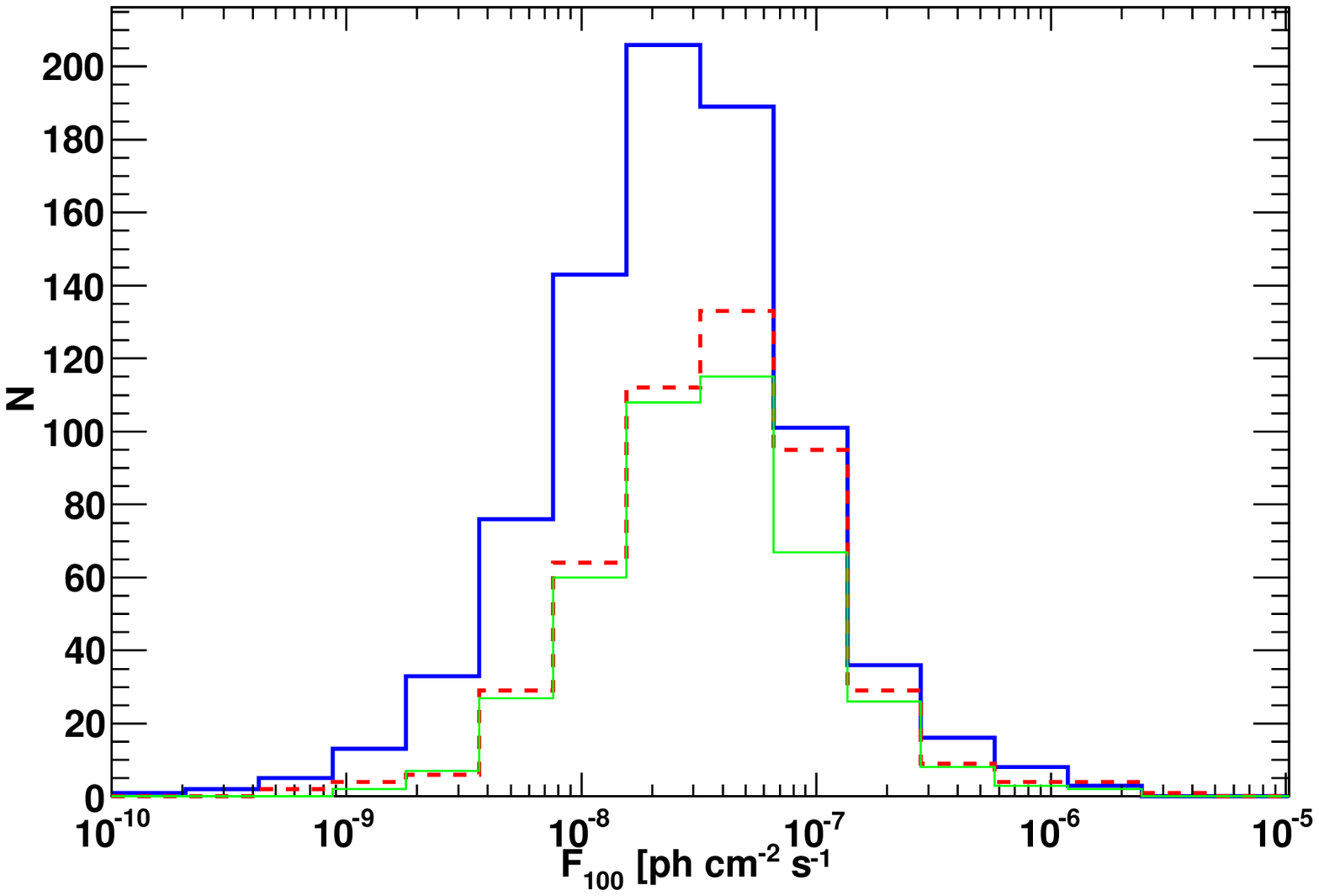} 
	\caption{Flux distributions of detected sources 
(TS$\geq$50 and $|b|\geq$20$^{\circ}$) for three different realizations
of the $\gamma$-ray sky. The solid thick line corresponds to a
log $N$ -- log $S$ distribution which resolves approximately $\sim$70\%
of the GeV diffuse background, while the dashed line corresponds to
the log $N$--log $S$ derived in this work which resolves approximately
$\sim$23\,\% of the diffuse background. For comparison the thin solid
line shows the flux distributions of the real sample of sources detected
by {\it Fermi}.
	\label{fig:flux_comp}}
\end{centering}
\end{figure}

In order to rule out the hypothesis that the sources detected by {\it Fermi} 
produce most
of the GeV diffuse emission, we performed an additional simulation.
In this exercise the input log $N$--log $S$ is compatible with
a single power law with a differential slope of 2.23. 
At bright fluxes this log $N$--log $S$ is compatible with
the one reported in \cite{lat_lbas} and at fluxes 
F$_{100}\geq10^{-9}$\,ph cm$^{-2}$ s$^{-1}$
accounts for $\sim$70\,\% of the EGB. In this scenario the surface
density of sources at  F$_{100}\geq10^{-9}$\,ph cm$^{-2}$ s$^{-1}$ is 
0.8\,deg$^{-2}$ (while the one we derived in $\S$~\ref{sec:comp} is
 0.12\,deg$^{-2}$).
To this simulation we applied
the same analysis steps used for both the real data and
the simulations analyzed in $\S$~\ref{sec:sim}.
Fig.~\ref{fig:flux_comp} compares  the flux distribution
of the sources detected in this simulation with the distribution
of the real sources
detected by LAT and also with the sources detected in
one of the simulations used in $\S$~\ref{sec:sim}.
It is apparent that the flux distribution of the sources
detected in the simulation under study here 
is very different from the other two. 

Indeed, in the case point-like sources produce most of the EGB
 {\it Fermi} should detect many more
medium-bright sources than are actually seen.
A Kolmogorv-Smirnov test yields that the probability that
the flux distribution (arising from the log $N$--log$S$ tested
in this section) comes from the same parent population as the real
data is $\leq10^{-5}$. This probability becomes $5\times10^{-4}$
if the $\chi^2$ test is used.
The KS test between the flux distribution of one of the simulations
used in $\S$~\ref{sec:sim} and the real data yields a probability of 
$\sim$87\,\% that both come from the same parent population while
it is $\sim$91\,\% if the $\chi^2$ test is used.

Thus the hypothesis that {\it Fermi} is resolving 
(for F$_{100}\geq10^{-9}$\,ph cm$^{-2}$ s$^{-1}$)
the majority
of the diffuse background can be ruled out at high confidence.

\section{Discussion and Conclusions}
\label{sec:discussion}

{\it Fermi} provides a huge leap in sensitivity for the study of the 
$\gamma$-ray sky with respect its predecessor EGRET. This work
focuses on the global intrinsic properties of the source
population detected by {\it Fermi} at high Galactic latitudes.

We constructed the source count distribution of all sources
detected above $|b|\geq$20$^{\circ}$. This distribution 
extends over three decades in flux and is compatible at
bright fluxes (e.g. F$_{100}\geq6\times10^{-8}$\,ph cm$^{-2}$ s$^{-1}$)
with a Euclidean function. Several methods have been employed to show
that at fainter fluxes the log $N$--log $S$ displays a significant
flattening.  We believe that this flattening has a cosmological
origin and is due to the fact that {\it Fermi} is already sampling,
with good accuracy,
the part of the luminosity function which shows negative evolution
(i.e. a decrease of the space density of sources with increasing
redshift). This is the first time that such flattening
has been found in the source count distributions of $\gamma$-ray sources
and blazars. We also showed that the log $N$--log $S$ of
blazars follows closely that of point source, showing that most
of the unassociated high-latitude sources in the 1FLG catalog
 are likely to be blazars. At the fluxes
currently sampled by {\it Fermi} (e.g. F$_{100}\geq10^{-9}$\,ph cm$^{-2}$ s$^{-1}$)
the surface density of blazars is 0.12$^{+0.03}_{-0.02}$\,deg$^{-2}$
and this is found to be a factor $\sim$4 larger than previous estimates.

The average intrinsic spectrum of blazars is in remarkably good agreement
with the spectrum of the GeV diffuse emission recently measured 
by {\it Fermi} \citep{lat_edb}. Nevertheless,
integrating the log $N$--log $S$, to the minimum detected source flux, 
shows that at least  16.0$^{+2.4}_{-2.6}$\,\% (the systematic
uncertainty is an additional 7\,\%) of the GeV background 
can be accounted for by  source populations measured by {\it Fermi}.
This is a small fraction of the total intensity and it is bound not to
increase dramatically 
unless the log $N$--log$S$ becomes steeper at fluxes below
$10^{-9}$\,ph cm$^{-2}$ s$^{-1}$. This generally never happens
unless a different source class starts to be detected in large
numbers at fainter fluxes.

\cite{thompson07} predict the integrated emission of starburst galaxies
to be $10^{-6}$\,ph cm$^{-2}$ s$^{-1}$ sr$^{-1}$ 
(above 100\,MeV). This would represent $\sim$10\,\%  of the LAT diffuse
background and would be comparable (although a bit less) to that 
of blazars found here. Indeed, their prediction that M82 and NGC 253 would
be the first two starburst galaxies to be detected has been fulfilled 
\citep{lat_starburst}. A similar contribution to the GeV diffuse background
should arise from the integrated emission of normal star forming galaxies
\citep{pavlidou02}. In both cases (normal and starburst galaxies) $\gamma$-rays
are produced from the interaction of cosmic rays with the interstellar gas
\cite[e.g. see][]{lat_cr}. It is natural to expect that
both normal and starburst galaxies produce a fraction of the diffuse emission
since now both classes are certified $\gamma$-ray sources \citep[see e.g.][]{cat1}.

It is also interesting to note that pulsars represent 
the second largest population
in our high-latitude sample (see Tab.~\ref{tab:sample}).
According to \cite{faucher09} pulsars and in particular
millisecond pulsars can produce a relevant fraction
of the GeV diffuse emission. However, given the strong break, typically
at  a few GeVs, in their spectra \citep[e.g. see][]{lat_vela2010}, 
millisecond pulsars  are not expected to contribute much
of the diffuse emission above a few GeVs.
Finally radio-quiet AGN might also contribute to the GeV diffuse background.
In these objects the $\gamma$-ray emission is supposedly produced
by a non-thermal electrons present in the corona
above the accretion disk \citep[see e.g.][for details]{inoue08}.
\cite{inoue09} predict that, at fluxes of
 F$_{100}\leq 10^{-10}$\,ph cm$^{-2}$ s$^{-1}$, radio-quiet AGN outnumber
the blazars.  According to their prediction, most of background could
be explained in terms of AGN (radio-quiet and radio-loud). 

It is thus clear that standard astrophysical scenarios can be invoked
to  explain the GeV extragalactic diffuse background. However,
the main result of this analysis is that blazars account only for $<$40\,\%
of it\footnote{This includes extrapolating the source counts
distribution to zero flux and taking into account statistical
and systematic uncertainties.}. It remains a mystery why the average spectrum
of blazars is so similar to the EGB spectrum. Taken by itself, this
finding would lead to believe that blazars might account for the entire
GeV diffuse background. However, we showed (see Fig.~\ref{fig:flux_comp} and 
$\S$~\ref{sec:siml} for details) that in this case {\it Fermi} should have
detected a much larger number (up to $\sim$50\,\%) of medium-bright
sources with a typical flux of  F$_{100}\geq10^{-8}$\,ph cm$^{-2}$ s$^{-1}$.
This scenario can thus be excluded with confidence.
Thus, the integrated emission from other source classes should still have
a spectrum declining as a power-law with an index of $\sim2.4$.
This does not seem to be a difficult problem to overcome.
Indeed, at least in the case of star forming galaxies 
we note that in the modeling
of both \cite{fields2010} and \cite{makiya2010} the integrated emission
from these sources displays a spectrum similar to the EGB one (at least
for energies above 200\,MeV).
Moreover, in this work  we also found  that the
 contribution to the diffuse emission of FSRQs and BL Lacs
is different, FSRQs being softer than BL Lacs. Thus, the summed
spectrum of their integrated diffuse emission is curved, softer
at low energy and hard at high ($>10$\,GeV) energy.
This makes it slightly different from the featureless power-law of the
diffuse background.
All the estimates presented here will be refined
with the derivation of the blazar luminosity
function which is left to a follow-up paper.

\clearpage
\acknowledgments
Helpful comments from the referee are acknowledged.
The \textit{Fermi} LAT Collaboration acknowledges generous ongoing support
from a number of agencies and institutes that have supported both the
development and the operation of the LAT as well as scientific data analysis.
These include the National Aeronautics and Space Administration and the
Department of Energy in the United States, the Commissariat \`a l'Energie Atomique
and the Centre National de la Recherche Scientifique / Institut National de Physique Nucl\'eaire et de Physique des Particules in France, the Agenzia 
Spaziale Italiana and the Istituto Nazionale di Fisica Nucleare in Italy, 
the Ministry of Education, Culture, Sports, Science and Technology (MEXT), 
High Energy Accelerator Research Organization (KEK) and Japan Aerospace 
Exploration Agency (JAXA) in Japan, and the K.~A.~Wallenberg Foundation, 
the Swedish Research Council and the Swedish National Space Board in Sweden.
Additional support for science analysis during the operations phase 
is gratefully acknowledged from the Istituto Nazionale di Astrofisica in 
Italy and the Centre National d'\'Etudes Spatiales in France.

{\it Facilities:} \facility{{\it Fermi}/LAT}.

\bibliographystyle{apj}
\bibliography{/Users/majello/Work/Papers/BiblioLib/biblio}

\end{document}